\documentclass[11pt,twoside,a4paper]{cernrep} 
\usepackage{rep_common}
\usepackage{cancel}
\usepackage{placeins}
\usepackage{booktabs}
\usepackage{slashbox}
\usepackage{subfig}
\usepackage{amsmath,amssymb,amsthm}
\usepackage{rotating}
\usepackage{multicol}
\usepackage{tabulary}
\usepackage[colorlinks, urlcolor=blue, linkcolor=blue, citecolor=blue, plainpages=false]{hyperref}
\usepackage{float}
\floatstyle{plaintop}
\restylefloat{table}
\usepackage{xargs}
\usepackage{xstring,ragged2e}

\usepackage{graphicx,multirow,colortbl}

\begin{document}
\newcounter{ncontacts}
\newcommand{\fcontact}[1]{\StrCount{#1}{,}[\tmp]\setcounter{ncontacts}{\tmp}
  Contact\ifthenelse{\value{ncontacts} > 0}{s}{}: #1}
\newcommand{\feditor}[1]{\StrCount{#1}{,}[\tmp]\setcounter{ncontacts}{\tmp}
  Contact Editor\ifthenelse{\value{ncontacts} > 0}{s}{}: #1}
\newcommandx{\asection}[2][1=NONE]{
  \ifthenelse{\equal{#1}{NONE}}
  {\section{#2}}{\section[#2]{#2\footnote{\feditor{#1}}}}}
\newcommandx{\asubsection}[2][1=NONE]{
  \ifthenelse{\equal{#1}{NONE}}
             {\section{#2}}{
               \section[#2]{#2\footnote{\fcontact{#1}}}}}
\newcommandx{\asubsubsection}[2][1=NONE]{
  \ifthenelse{\equal{#1}{NONE}}
             {\subsubsection{#2}}{
               \subsubsection[#2]{#2\footnote{\fcontact{#1}}}}}
\newcommandx{\asubsubsubsection}[2][1=NONE]{
  \ifthenelse{\equal{#1}{NONE}}
             {\subsubsubsection{#2}}{
               \subsubsubsection[#2]{#2\footnote{\fcontact{#1}}}}}
               
\newcommand\snowmass{\begin{center}\rule[-0.2in]{\hsize}{0.01in}\\\rule{\hsize}{0.01in}\\
\vskip 0.1in Submitted to the  Proceedings of the US Community Study\\ 
on the Future of Particle Physics (Snowmass 2021)\\ 
\rule{\hsize}{0.01in}\\\rule[+0.2in]{\hsize}{0.01in} \end{center}}


\title{{\normalfont\bfseries\boldmath\huge
\begin{center}
Muon Collider Physics Summary
\end{center}
\vspace*{-35pt}
}
{\textnormal{\normalsize \snowmass
\vspace*{-40pt}
}}
{\textnormal{\normalsize
\abstract{
The perspective of designing muon colliders with high energy and luminosity, which is being investigated by the International Muon Collider Collaboration, has triggered a growing interest in their physics reach. 
\ \\[2pt]
We present a concise summary of the muon collider potential to explore new physics, leveraging on the unique possibility of combining high available en- ergy with very precise measurements.  
}\\[30pt]}}
{\textnormal{\normalsize\justifying
This is one of the five reports submitted to Snowmass by the muon colliders community at large. The reports preparation effort has been coordinated by the International Muon Collider Collaboration. Authors and Signatories have been collected with a 
\href{https://indico.cern.ch/event/1130036/}{subscription page}, and are defined as follows:
\begin{itemize}
 \item An ``Author'' contributed to the results documented in the report in any form, including e.g.~by participating to the discussions of the community meetings and sending comments on the draft, or plans to contribute to the future work.
\item
A ``Signatory'' expresses support to the efforts described in the report and endorses the Collaboration plans.
\end{itemize}
}}
}
\newcounter{instituteref}
\newcommand{\iinstitute}[2]{\refstepcounter{instituteref}\label{#1}$^{\ref{#1}}$\href{http://inspirehep.net/record/#1}{#2}}
\newcommand{\iauthor}[3]{\href{http://inspirehep.net/record/#1}{#2}$^{#3}$}
\author{Authors: \\
    \iauthor{1757334}{C.~Aim\`e}{\ref{943385},\ref{902885}},
    \iauthor{1060487}{A.~Apyan}{\ref{902682}},
    \iauthor{1067349}{M.A.~Mahmoud.}{\ref{912409}},
    \iauthor{1073143}{N.~Bartosik}{\ref{902889}},
    \iauthor{1029828}{A.~Bertolin}{\ref{902884}},
    \iauthor{1015872}{M.~Bonesini}{\ref{902882},\ref{907960}},
    \iauthor{1794682}{S.~Bottaro}{\ref{903128},\ref{902886}},
    \iauthor{1077579}{D.~Buttazzo}{\ref{902886}},
    \iauthor{1275234}{R.~Capdevilla}{\ref{908474},\ref{903282}},
    \iauthor{1057458}{M.~Casarsa}{\ref{902888}},
    \iauthor{}{L.~Castelli}{\ref{903113}},
    \iauthor{1014281}{M.G.~Catanesi}{\ref{902877}},
    \iauthor{1418744}{F.G.~Celiberto}{\ref{906718},\ref{912328}},
    \iauthor{1021757}{A.~Cerri}{\ref{1241166}},
    \iauthor{1793525}{C.~Cesarotti}{\ref{902835}},
    \iauthor{1037833}{G.~Chachamis}{\ref{905303}},
    \iauthor{2037614}{S.~Chen}{\ref{1471035}},
    \iauthor{1069708}{Y.-T.~Chien}{\ref{1275736}},
    \iauthor{1272180}{M.~Chiesa}{\ref{943385},\ref{902885}},
    \iauthor{1013241}{G.~Collazuol}{\ref{902884},\ref{903113}},
    \iauthor{1862239}{M.~Costa}{\ref{903128},\ref{902886}},
    \iauthor{1046385}{N.~Craig}{\ref{903307}},
    \iauthor{1024481}{D.~Curtin}{\ref{903282}},
    \iauthor{1012395}{S.~Dasu}{\ref{903349}},
    \iauthor{1021004}{J.~de~Blas}{\ref{903836}},
    \iauthor{1012030}{D.~Denisov}{\ref{902689}},
    \iauthor{1012025}{H.~Denizli}{\ref{908452}},
    \iauthor{1011983}{R.~Dermisek}{\ref{902874}},
    \iauthor{1063643}{L.~Luzio}{\ref{903113},\ref{902884}},
    \iauthor{1031269}{B.~Di~Micco}{\ref{906528},\ref{907692}},
    \iauthor{1011804}{K.~R.~Dienes}{\ref{902647},\ref{902990}},
    \iauthor{1011508}{T.~Dorigo}{\ref{902884}},
    \iauthor{1010105}{A.~Ferrari}{\ref{1276460}},
    \iauthor{1719039}{D.~Fiorina}{\ref{902885}},
    \iauthor{1052115}{R.~Franceschini}{\ref{906528},\ref{907692}},
    \iauthor{1946817}{F.~Garosi}{\ref{904416}},
    \iauthor{1706734}{A.~Glioti}{\ref{1471035}},
    \iauthor{}{M.~Greco}{\ref{906528}},
    \iauthor{1198373}{A.~Greljo}{\ref{902668}},
    \iauthor{1258537}{R.~Gr\"ober}{\ref{1513358},\ref{902884}},
    \iauthor{1007486}{C.~Grojean}{\ref{902770},\ref{902858}},
    \iauthor{1274618}{J.~Gu}{\ref{903628}},
    \iauthor{1006825}{T.~Han}{\ref{903130}},
    \iauthor{1383268}{B.~Henning}{\ref{1471035}},
    \iauthor{1912097}{K.~Hermanek}{\ref{902874}},
    \iauthor{1067690}{T.R.~Holmes}{\ref{1623978}},
    \iauthor{1515880}{S.~Homiller}{\ref{902835}},
    \iauthor{1475406}{S.~Jana}{\ref{902841}},
    \iauthor{1028687}{S.~Jindariani}{\ref{902796}},
    \iauthor{1051663}{Y.~Kahn}{\ref{902867}},
    \iauthor{1653554}{I.~Karpov}{\ref{902725}},
    \iauthor{1002991}{W.~Kilian}{\ref{903203}},
    \iauthor{1020007}{K.~Kong}{\ref{902912}},
    \iauthor{1019544}{P.~Koppenburg}{\ref{903832}},
    \iauthor{1252769}{K.~Krizka}{\ref{902953}},
    \iauthor{1071846}{L.~Lee}{\ref{1623978}},
    \iauthor{1074984}{Q.~Li}{\ref{903603}},
    \iauthor{1000076}{R.~Lipton}{\ref{902796}},
    \iauthor{1256188}{Z.~Liu}{\ref{903010}},
    \iauthor{999862}{K.R.~Long}{\ref{902868},\ref{903174}},
    \iauthor{999724}{I.~Low}{\ref{902645},\ref{903083}},
    \iauthor{999654}{D.~Lucchesi}{\ref{903113},\ref{902884}},
    \iauthor{1514492}{Y.~Ma}{\ref{903130}},
    \iauthor{1355155}{L.~Ma}{\ref{904187}},
    \iauthor{999053}{F.~Maltoni}{\ref{910783},\ref{902674}},
    \iauthor{998923}{B.~Mansouli\'e}{\ref{912490}},
    \iauthor{1668860}{L.~Mantani}{\ref{907623}},
    \iauthor{1078065}{D.~Marzocca}{\ref{902888}},
    \iauthor{1751811}{N.~McGinnis}{\ref{903290}},
    \iauthor{997877}{B.~Mele}{\ref{902887}},
    \iauthor{1074063}{F.~Meloni}{\ref{902770}},
    \iauthor{1461119}{C.~Merlassino}{\ref{903112}},
    \iauthor{2049482}{A.~Montella}{\ref{902888}},
    \iauthor{1069385}{M.~Nardecchia}{\ref{903168},\ref{902887}},
    \iauthor{}{F.~Nardi}{\ref{903113},\ref{902884}},
    \iauthor{1077958}{P.~Panci}{\ref{903129},\ref{902886}},
    \iauthor{1048820}{S.~Pagan~Griso}{\ref{902953}},
    \iauthor{1037621}{G.~Panico}{\ref{902801},\ref{902880}},
    \iauthor{1067971}{R.~Paparella}{\ref{907142}},
    \iauthor{1023838}{P.~Paradisi}{\ref{1513358},\ref{902884}},
    \iauthor{994095}{N.~Pastrone}{\ref{902889}},
    \iauthor{993440}{F.~Piccinini}{\ref{902885}},
    \iauthor{1050691}{K.~Potamianos}{\ref{903112}},
    \iauthor{992463}{E.~Radicioni}{\ref{902877}},
    \iauthor{992222}{R.~Rattazzi}{\ref{1471035}},
    \iauthor{1214912}{D.~Redigolo}{\ref{902880}},
    \iauthor{992031}{L.~Reina}{\ref{902803}},
    \iauthor{1021811}{J.~Reuter}{\ref{902770}},
    \iauthor{1020819}{C.~Riccardi}{\ref{943385},\ref{902885}},
    \iauthor{1885919}{L.~Ricci}{\ref{1471035}},
    \iauthor{1028713}{L.~Ristori}{\ref{902796}},
    \iauthor{1040385}{T.~Robens}{\ref{902678}},
    \iauthor{1054727}{R.~Ruiz}{\ref{902756}},
    \iauthor{1072232}{F.~Sala}{\ref{908583}},
    \iauthor{1885424}{J.~Salko}{\ref{902668}},
    \iauthor{990505}{P.~Salvini}{\ref{902885}},
    \iauthor{1077871}{E.~Salvioni}{\ref{1513358},\ref{902884}},
    \iauthor{989645}{D.~Schulte}{\ref{902725}},
    \iauthor{1039590}{M.~Selvaggi}{\ref{902725}},
    \iauthor{1019799}{A.~Senol}{\ref{908452}},
    \iauthor{1342183}{L.~Sestini}{\ref{902884}},
    \iauthor{1071696}{V.~Sharma}{\ref{903349}},
    \iauthor{1019902}{J.~Shu}{\ref{903895}},
    \iauthor{1074094}{R.~Simoniello}{\ref{902725}},
    \iauthor{1319078}{G.~Stark}{\ref{1218068}},
    \iauthor{1023964}{D.~Stolarski}{\ref{906105}},
    \iauthor{987285}{S.~Su}{\ref{902647}},
    \iauthor{1027860}{W.~Su}{\ref{907284}},
    \iauthor{1375692}{O.~Sumensari}{\ref{1776405}},
    \iauthor{1071725}{X.~Sun}{\ref{1210798}},
    \iauthor{987128}{R.~Sundrum}{\ref{912511}},
    \iauthor{1268914}{J.~Tang}{\ref{903702},\ref{903123}},
    \iauthor{1257546}{A.~Tesi}{\ref{902880}},
    \iauthor{1023463}{B.~Thomas}{\ref{904505}},
    \iauthor{1064514}{R.~Torre}{\ref{902881}},
    \iauthor{1778841}{S.~Trifinopoulos}{\ref{902888}},
    \iauthor{1265350}{I.~Vai}{\ref{902885}},
    \iauthor{1880878}{A.~Valenti}{\ref{1513358},\ref{902884}},
    \iauthor{1863232}{L.~Vittorio}{\ref{903128},\ref{902886}},
    \iauthor{984146}{L.-T.~Wang}{\ref{902729}},
    \iauthor{1511975}{Y.~Wu}{\ref{903094}},
    \iauthor{1037622}{A.~Wulzer}{\ref{903113}},
    \iauthor{1656814}{X.~Zhao}{\ref{906528},\ref{907692}},
    \iauthor{1037623}{J.~Zurita}{\ref{907907}}
    \\ \vspace*{4mm} Signatories: \\
    \iauthor{1018987}{D.~Acosta}{\ref{903156}},
    \iauthor{1018902}{K.~Agashe}{\ref{902990}},
    \iauthor{1018633}{B.C.~Allanach}{\ref{907623}},
    \iauthor{1018264}{F.~Anulli}{\ref{902887}},
    \iauthor{1049113}{A.~Apresyan}{\ref{902796}},
    \iauthor{1491320}{P.~Asadi}{\ref{1237813}},
    \iauthor{}{D.~Athanasakos}{\ref{910429}},
    \iauthor{1041900}{A.~Azatov}{\ref{904416},\ref{902888}},
    \iauthor{1028433}{J.J.~Back}{\ref{903734}},
    \iauthor{1424044}{L.~Bandiera}{\ref{905268}},
    \iauthor{1017330}{R.~J.~Barlow}{\ref{911708}},
    \iauthor{1037853}{E.~Barzi}{\ref{902796},\ref{903092}},
    \iauthor{2031609}{F.~Batsch}{\ref{902725}},
    \iauthor{1068289}{M.~Bauce}{\ref{902887},\ref{903168}},
    \iauthor{1016672}{J.~S.~Berg}{\ref{902689}},
    \iauthor{1016557}{J.~Berryhill}{\ref{902796}},
    \iauthor{1049763}{A.~Bersani}{\ref{902881}},
    \iauthor{1020223}{K.M.~Black}{\ref{903349}},
    \iauthor{1015812}{C.~Booth}{\ref{903196}},
    \iauthor{}{L.~Bottura}{},
    \iauthor{1031261}{D.~Bowring}{\ref{902796}},
    \iauthor{1015478}{A.~Braghieri}{\ref{902885}},
    \iauthor{1015228}{G.~Brooijmans}{\ref{902749}},
    \iauthor{1015214}{A.~Bross}{\ref{902796}},
    \iauthor{1114205}{E.~Brost}{\ref{902689}},
    \iauthor{1894439}{L.~Buonincontri}{\ref{902884},\ref{903113}},
    \iauthor{1670912}{B.~Caiffi}{\ref{902881}},
    \iauthor{1014742}{G.~Calderini}{\ref{926589},\ref{903119}},
    \iauthor{1707397}{S.~Calzaferri}{\ref{902885}},
    \iauthor{}{P.~Cameron}{\ref{902689}},
    \iauthor{1024602}{A.~Canepa}{\ref{902796}},
    \iauthor{}{F.~Casaburo}{},
    \iauthor{1020772}{G.~Cavoto}{\ref{903168},\ref{902887}},
    \iauthor{1075318}{L.~Celona}{\ref{902879}},
    \iauthor{}{G.~Cesarini}{},
    \iauthor{1014143}{Z.~Chacko}{\ref{902990}},
    \iauthor{2023221}{A.~Chanc\'e}{\ref{912490}},
    \iauthor{1380376}{R.~T.~Co}{\ref{908012}},
    \iauthor{1013275}{A.~Colaleo}{\ref{902660},\ref{902877}},
    \iauthor{}{D.~J.~Colling}{\ref{902868}},
    \iauthor{1013050}{G.~Corcella}{\ref{902807}},
    \iauthor{}{L.~M.~Cremaldi}{},
    \iauthor{1060042}{A.~Crivellin}{\ref{903370},\ref{905405}},
    \iauthor{1035631}{Y.~Cui}{\ref{903304}},
    \iauthor{1937290}{C.~Curatolo}{\ref{902882}},
    \iauthor{1067364}{R.~T.~D'Agnolo}{\ref{1087875}},
    \iauthor{1047966}{F.~D'Eramo}{\ref{903113},\ref{902884}},
    \iauthor{2049478}{G.~Da~Molin}{\ref{903113}},
    \iauthor{2052018}{M.~Dam}{\ref{902882}},
    \iauthor{1076225}{H.~Damerau}{\ref{902725}},
    \iauthor{1651018}{E.~De~Matteis}{\ref{907142}},
    \iauthor{1012237}{A.~Deandrea}{\ref{1743848}},
    \iauthor{1012143}{J.~Delahaye}{\ref{902725}},
    \iauthor{1019723}{A.~Delgado}{\ref{903085}},
    \iauthor{1039487}{C.~Densham}{\ref{903174}},
    \iauthor{1395010}{K.~F.~Di~Petrillo}{\ref{902796}},
    \iauthor{1246709}{J.~Dickinson}{\ref{902796}},
    \iauthor{1011443}{M.~Dracos}{\ref{911366}},
    \iauthor{1054778}{J.~Duarte}{\ref{903305}},
    \iauthor{1404358}{F.~Errico}{\ref{902660},\ref{902877}},
    \iauthor{1048347}{R.~Essig}{\ref{910429}},
    \iauthor{1064320}{P.~Everaerts}{\ref{903349}},
    \iauthor{1010523}{L.~Everett}{\ref{903349}},
    \iauthor{1010482}{M.~Fabbrichesi}{\ref{902888}},
    \iauthor{1045844}{J.~Fan}{\ref{902692}},
    \iauthor{1069878}{S.~Farinon}{\ref{902881}},
    \iauthor{1648215}{J.~F.~Somoza}{\ref{902725}},
    \iauthor{1010065}{G.~Ferretti}{\ref{902825}},
    \iauthor{1009979}{F.~Filthaut}{\ref{903075}},
    \iauthor{1894571}{M.~Forslund}{\ref{910429}},
    \iauthor{1046463}{P.~Franchini}{\ref{902948},\ref{903170}},
    \iauthor{1019509}{M.~Frigerio}{\ref{1508424}},
    \iauthor{1009120}{E.~Gabrielli}{\ref{903287},\ref{902888}},
    \iauthor{1009009}{M.~Gallinaro}{\ref{905303}},
    \iauthor{1068164}{I.~Garcia~Garcia}{\ref{903889}},
    \iauthor{1894454}{L.~Giambastiani}{\ref{903113},\ref{902884}},
    \iauthor{}{A.S.~Giannakopoulou}{\ref{903237}},
    \iauthor{1075917}{D.~Giove}{\ref{907142}},
    \iauthor{1971617}{C.~Giraldin}{\ref{903113}},
    \iauthor{1029806}{L.~Gladilin}{},
    \iauthor{1008112}{S.~Goldfarb}{\ref{902999}},
    \iauthor{1037882}{H.M.~Gray}{\ref{903299},\ref{902953}},
    \iauthor{1059457}{L.~Gray}{\ref{902796}},
    \iauthor{1007092}{H.E.~Haber}{\ref{1218068}},
    \iauthor{1055424}{J.~Haley}{\ref{903094}},
    \iauthor{1259916}{C.~Han}{\ref{903702}},
    \iauthor{2044726}{J.~Hauptman}{\ref{902893}},
    \iauthor{1006149}{M.~Herndon}{\ref{903349}},
    \iauthor{2049476}{H.~Jia}{\ref{903349}},
    \iauthor{}{C.~Jolly}{\ref{903174}},
    \iauthor{1003695}{D.~M.~Kaplan}{\ref{902865}},
    \iauthor{1067609}{D.~Kelliher}{\ref{903174}},
    \iauthor{1345391}{G.K.~Krintiras}{\ref{902912}},
    \iauthor{1077491}{G.~Krnjaic}{\ref{902796}},
    \iauthor{1854911}{N.~Kumar}{\ref{902767}},
    \iauthor{1001375}{P.~Kyberd}{\ref{903940}},
    \iauthor{999784}{R.~LOSITO}{\ref{902725}},
    \iauthor{1292423}{J.-B.~Lagrange}{\ref{903174}},
    \iauthor{}{S.~Levorato}{},
    \iauthor{1064657}{W.~Li}{\ref{903156}},
    \iauthor{1996476}{R.~L.~Voti}{\ref{902887}},
    \iauthor{1393097}{D.~Liu}{\ref{1424710}},
    \iauthor{1074693}{M.~Liu}{\ref{903142}},
    \iauthor{}{S.~Lomte}{\ref{903349}},
    \iauthor{1700371}{Q.~Lu}{\ref{902835}},
    \iauthor{1041997}{R.~Mahbubani}{\ref{902678}},
    \iauthor{1056868}{A.~Mariotti}{\ref{907933}},
    \iauthor{1670119}{S.~Mariotto}{\ref{903009},\ref{907142}},
    \iauthor{1971307}{P.~Mastrapasqua}{\ref{910783}},
    \iauthor{998430}{K.~Matchev}{\ref{902804}},
    \iauthor{1054925}{A.~Mazzacane}{\ref{902796}},
    \iauthor{1025277}{P.~Meade}{\ref{910429}},
    \iauthor{1022138}{P.~Merkel}{\ref{902796}},
    \iauthor{1032624}{F.~Mescia}{\ref{905190},\ref{911212}},
    \iauthor{1066275}{R.~K.~Mishra}{\ref{902835}},
    \iauthor{1070072}{A.~Mohammadi}{\ref{903349}},
    \iauthor{}{R.~Mohapatra}{},
    \iauthor{997065}{N.~Mokhov}{\ref{902796}},
    \iauthor{996989}{P.~Montagna}{\ref{943385},\ref{902885}},
    \iauthor{1064125}{R.~Musenich}{\ref{902881}},
    \iauthor{995835}{M.S.~Neubauer}{\ref{902867}},
    \iauthor{995826}{D.~Neuffer}{\ref{902796}},
    \iauthor{995794}{H.~Newman}{\ref{902711}},
    \iauthor{995460}{Y.~Nomura}{\ref{903299}},
    \iauthor{1070110}{I.~Ojalvo}{\ref{16750}},
    \iauthor{1465792}{J.L.~Oliver}{\ref{903302}},
    \iauthor{1063052}{G.~Ortona}{\ref{902889}},
    \iauthor{1274353}{D.~Pagani}{\ref{902878}},
    \iauthor{994435}{M.~Palmer}{\ref{902689}},
    \iauthor{1772198}{A.~Pellecchia}{\ref{902660}},
    \iauthor{1067962}{A.~Perloff}{\ref{902748}},
    \iauthor{1021028}{M.~Pierini}{\ref{902725}},
    \iauthor{1651162}{M.~Prioli}{\ref{907142}},
    \iauthor{1024769}{M.~Procura}{\ref{903326}},
    \iauthor{1217056}{R.~Radogna}{\ref{902660},\ref{902877}},
    \iauthor{991702}{R.A.~Rimmer}{\ref{904961}},
    \iauthor{1056642}{F.~Riva}{\ref{902813}},
    \iauthor{1054170}{C.~Rogers}{\ref{903174}},
    \iauthor{991185}{L.~Rossi}{\ref{903009},\ref{907142}},
    \iauthor{1912150}{R.~Ryne}{\ref{1189711}},
    \iauthor{990367}{J.~Santiago}{\ref{909079},\ref{903836}},
    \iauthor{1042797}{E.~Santopinto}{\ref{902881}},
    \iauthor{}{I.~Sarra}{},
    \iauthor{989950}{J.~Schieck}{\ref{903324},\ref{904536}},
    \iauthor{1022121}{R.~Schwiehorst}{\ref{903006}},
    \iauthor{1055618}{D.~Sertore}{\ref{907142}},
    \iauthor{988978}{V.~Shiltsev}{\ref{902796}},
    \iauthor{988660}{L.~Silvestris}{\ref{902877}},
    \iauthor{1622677}{F.~M.~Simone}{\ref{902660},\ref{902877}},
    \iauthor{1889775}{K.~Skoufaris}{\ref{902725}},
    \iauthor{1046340}{P.~Snopok}{\ref{902865}},
    \iauthor{988143}{F.J.P.~Soler}{\ref{902823}},
    \iauthor{1066476}{M.~Sorbi}{\ref{903009},\ref{907142}},
    \iauthor{}{A.~Stamerra}{\ref{902660},\ref{902877}},
    \iauthor{1057643}{M.~Statera}{\ref{907142}},
    \iauthor{1058749}{D.~Stratakis}{\ref{902796}},
    \iauthor{1077402}{N.~Strobbe}{\ref{903010}},
    \iauthor{1071880}{J.~Stupak}{\ref{1273509}},
    \iauthor{1078570}{M.~Swiatlowski}{\ref{903290}},
    \iauthor{1454316}{A.~Sytov}{\ref{905268}},
    \iauthor{1020371}{A.~Taffard}{\ref{903302}},
    \iauthor{986860}{T.~Tait}{\ref{903302}},
    \iauthor{2057899}{J.~Tang}{\ref{903123}},
    \iauthor{1055960}{M.~Taoso}{\ref{902889}},
    \iauthor{1054400}{J.~Thaler}{\ref{1237813}},
    \iauthor{1878399}{E.~A.~Thompson}{\ref{902770}},
    \iauthor{985810}{L.~Tortora}{\ref{907692}},
    \iauthor{1031757}{Y.~Torun}{\ref{902865}},
    \iauthor{1613622}{M.~Valente}{\ref{903290}},
    \iauthor{2025179}{R.~U.~Valente}{\ref{907142}},
    \iauthor{1643523}{N.~Valle}{\ref{943385},\ref{902885}},
    \iauthor{1071756}{R.~Venditti}{\ref{902660},\ref{902877}},
    \iauthor{1063935}{P.~Verwilligen}{\ref{902877}},
    \iauthor{1077738}{N.~Vignaroli}{\ref{902883}},
    \iauthor{984555}{P.~Vitulo}{\ref{943385},\ref{902885}},
    \iauthor{1077733}{E.~Vryonidou}{\ref{902984}},
    \iauthor{1054127}{C.~Vuosalo}{\ref{903349}},
    \iauthor{1073818}{H.~Weber}{\ref{902858}},
    \iauthor{1260509}{C.G.~Whyte}{\ref{904214}},
    \iauthor{1618109}{K.~Xie}{\ref{903130}},
    \iauthor{982905}{A.~Yamamoto}{\ref{902916}},
    \iauthor{1421892}{W.~Yin}{\ref{903268}},
    \iauthor{1019845}{K.~Yonehara}{\ref{902796}},
    \iauthor{1024759}{H.-B.~Yu}{\ref{903304}},
    \iauthor{1064691}{M.~Zanetti}{\ref{903113}},
    \iauthor{1971310}{A.~Zaza}{\ref{902660},\ref{902877}},
    \iauthor{}{J.~Zhang}{},
    \iauthor{1066114}{Y.~J.~Zheng}{\ref{902912}},
    \iauthor{981974}{A.~Zlobin}{\ref{902796}},
    \iauthor{1863481}{D.~Zuliani}{\ref{903113},\ref{902884}}
    \vspace*{1cm}} \institute{\small
\iinstitute{943385}{Universit{\`a} di Pavia, Italy};
\iinstitute{902682}{Department of Physics, Brandeis University, United States};
\iinstitute{912409}{{Center for High Energy Physics (CHEP-FU), Fayoum University, Egypt}};
\iinstitute{902889}{{INFN Sezione di Torino, Italy}};
\iinstitute{902884}{INFN Sezione di Padova, Italy};
\iinstitute{902882}{Istituto Nazionale di Fisica Nucleare, Italy};
\iinstitute{903128}{{Scuola Normale Superiore, Italy}};
\iinstitute{902886}{INFN Sezione di Pisa, Italy};
\iinstitute{908474}{Perimeter Institute, Canada};
\iinstitute{902888}{INFN Sezione di Trieste, Italy};
\iinstitute{903113}{Dipartimento di Fisica e Astronomia, Universit'a di Padova, Italy};
\iinstitute{902877}{INFN Sezione di Bari, Italy};
\iinstitute{906718}{European Centre for Theoretical Studies in Nuclear Physics and Related Areas (ECT*), Italy};
\iinstitute{1241166}{MPS School, University of Sussex, United Kingdom};
\iinstitute{902835}{Department of Physics, Harvard University, United States};
\iinstitute{905303}{Laborat{\' o}rio de Instrumenta\c{c}{\~ a}o e F{\' \i}sica Experimental de Part{\' \i}culas (LIP), Portugal};
\iinstitute{1471035}{{Theoretical Particle Physics Laboratory (LPTP), Institute of Physics, EPFL, Switzerland}};
\iinstitute{1275736}{Physics and Astronomy Department, Georgia State University, United States};
\iinstitute{903307}{University of California, Santa Barbara, United States};
\iinstitute{903282}{Department of Physics, University of Toronto, Canada};
\iinstitute{903349}{University of Wisconsin, United States};
\iinstitute{903836}{CAFPE and Departamento de F\'isica Te\'orica y del Cosmos, Universidad de Granada, Spain};
\iinstitute{902689}{Brookhaven National Laboratory, United States};
\iinstitute{908452}{Department of Physics, Bolu Abant Izzet Baysal University, Turkey};
\iinstitute{902874}{Physics Department, Indiana University, United States};
\iinstitute{906528}{Dipartimento di Matematica e Fisica, Universit\`a Roma Tre, Italy};
\iinstitute{902647}{Department of Physics, University of Arizona, United States};
\iinstitute{1276460}{Helmholtz-Zentrum Dresden-Rossendorf, Germany};
\iinstitute{902885}{INFN, Sezione di Pavia, Italy};
\iinstitute{904416}{SISSA, Italy};
\iinstitute{902668}{Albert Einstein Center for Fundamental Physics, Institute for Theoretical Physics, University of Bern, Switzerland};
\iinstitute{1513358}{Universit\`a di Padova, Italy};
\iinstitute{902770}{{Deutsches Elektronen-Synchrotron DESY, Germany}};
\iinstitute{903628}{Department of Physics, Fudan University, China};
\iinstitute{903130}{University of Pittsburgh, United States};
\iinstitute{1623978}{University of Tennessee, United States};
\iinstitute{902841}{Max-Planck-Institut f{\"u}r Kernphysik, Germany};
\iinstitute{902796}{Fermi National Accelerator Laboratory, United States};
\iinstitute{902867}{Department of Physics, University of Illinois at Urbana-Champaign, United States};
\iinstitute{902725}{CERN, Switzerland};
\iinstitute{903203}{Department of Physics, University of Siegen, Germany};
\iinstitute{902912}{Department of Physics and Astronomy, University of Kansas, United States};
\iinstitute{903832}{Nikhef National Institute for Subatomic Physics, The Netherlands};
\iinstitute{902953}{{Physics Division, Lawrence Berkeley National Laboratory, United States}};
\iinstitute{903603}{Peking University, China};
\iinstitute{903010}{School of Physics and Astronomy, University of Minnesota, United States};
\iinstitute{902868}{Imperial College London, United Kingdom};
\iinstitute{902645}{High Energy Physics Division, Argonne National Laboratory, United States};
\iinstitute{904187}{Shandong University, China};
\iinstitute{910783}{Center for Cosmology, Particle Physics and Phenomenology, Universit\'e catholique de Louvain, Belgium};
\iinstitute{912490}{IRFU, CEA, University Paris-Saclay, France};
\iinstitute{907623}{{DAMTP, University of Cambridge, United Kingdom}};
\iinstitute{903290}{TRIUMF, Canada};
\iinstitute{902887}{{INFN Sezione di Roma, Italy}};
\iinstitute{903112}{Particle Physics Department, University of Oxford, United Kingdom};
\iinstitute{903168}{Sapienza University of Rome, Italy};
\iinstitute{903129}{Pisa University, Italy};
\iinstitute{902801}{{Dipartimento di Fisica e Astronomia, Universit{\`a} degli Studi di Firenze, Italy}};
\iinstitute{907142}{INFN Sezione di Milano, LASA, Italy};
\iinstitute{902880}{INFN Sezione di Firenze, Italy};
\iinstitute{902803}{Florida State University, United States};
\iinstitute{902678}{Rudjer Boskovic Institute, Croatia};
\iinstitute{902756}{Institute of Nuclear Physics -- Polish Academy of Sciences {\rm (IFJ PAN)}, Poland};
\iinstitute{908583}{Laboratoire de Physique Th\'eorique et Hautes \'Energies, Sorbonne Universit\'e, CNRS, France};
\iinstitute{903895}{CAS Key Laboratory of Theoretical Physics, Insitute of Theoretical Physics, Chinese Academy of Sciences, P.R.China};
\iinstitute{1218068}{SCIPP, UC Santa Cruz, United States};
\iinstitute{906105}{Ottawa-Carleton Institute for Physics, Carleton University, Canada};
\iinstitute{907284}{Korea Institute for Advanced Study, South Korea};
\iinstitute{1776405}{IJCLab, P\^ole Th\'eorie (B\^at.~210), CNRS/IN2P3 et Universit\'e Paris-Saclay, France};
\iinstitute{1210798}{State Key Laboratory of Nuclear Physics and Technology, Peking University, China};
\iinstitute{912511}{Maryland Center for Fundamental Physics, University of Maryland, United States};
\iinstitute{903702}{Sun Yat-sen University, China};
\iinstitute{904505}{Department of Physics, Lafayette College, United States};
\iinstitute{902881}{INFN Sezione di Genova, Italy};
\iinstitute{902729}{Department of Physics, University of Chicago, United States};
\iinstitute{903094}{Department of Physics, Oklahoma State University, United States};
\iinstitute{907907}{{Instituto de F{\'i}sica Corpuscular, CSIC-Universitat de Val{\'e}ncia, Spain}};
\iinstitute{1111512}{Institut f{\"u}r Allgemeine Elektrotechnik, Universit{\"a}t Rostock, Germany};
\iinstitute{907960}{Dipartimento di Fisica, Universit\`a Milano Bicocca, Italy};
\iinstitute{912328}{INFN-TIFPA Trento Institute of Fundamental Physics and Applications, Italy};
\iinstitute{907692}{INFN Sezione di Roma Tre, Italy};
\iinstitute{902990}{Department of Physics, University of Maryland, United States};
\iinstitute{902858}{{Humboldt-Universit\"at zu Berlin, Institut f\"ur Physik, Germany}};
\iinstitute{903174}{STFC, United Kingdom};
\iinstitute{903083}{Department of Physics and Astronomy, Northwestern University, United States};
\iinstitute{902674}{Dipartimento di Fisica e Astronomia, Universit\`a di Bologna, Italy};
\iinstitute{903123}{Institute of High-Energy Physics, China};
\iinstitute{903156}{Physics \& Astronomy Department, Rice University, United States};
\iinstitute{1237813}{Center for Theoretical Physics, Massachusetts Institute of Technology, United States};
\iinstitute{910429}{YITP, Stony Brook, United States};
\iinstitute{903734}{Department of Physics, University of Warwick, United Kingdom};
\iinstitute{905268}{INFN Sezione di Ferrara, Italy};
\iinstitute{911708}{The University of Huddersfield, United Kingdom};
\iinstitute{903196}{{Department of Physics and Astronomy, University of Sheffield, United Kingdom}};
\iinstitute{902749}{Columbia University, United States};
\iinstitute{926589}{CNRS/IN2P3, France};
\iinstitute{902879}{INFN Sezione di Catania, Italy};
\iinstitute{908012}{University of Minnesota, United States};
\iinstitute{902660}{{Department of Physics, Universit{\`a} degli Studi di Bari, Italy}};
\iinstitute{902807}{INFN, Laboratori Nazionali di Frascati, Italy};
\iinstitute{903370}{University of Zurich, Switzerland};
\iinstitute{903304}{University of California-Riverside, United States};
\iinstitute{1087875}{Universit\`e Paris Saclay, CNRS, CEA, Institut de Physique Th\`eorique, France};
\iinstitute{1743848}{IP2I, Universit\'e Lyon 1, CNRS/IN2P3, France};
\iinstitute{903085}{University of Notre Dame, United States};
\iinstitute{911366}{IPHC, Universit\'{e} de Strasbourg, CNRS/IN2P3, France};
\iinstitute{903305}{University of California San Diego, United States};
\iinstitute{902692}{Brown University, United States};
\iinstitute{902825}{Chalmers University of Technology, Sweden};
\iinstitute{903075}{Radboud University and Nikhef, The Netherlands};
\iinstitute{902948}{{University of Lancaster, Department of Physics, United Kingdom}};
\iinstitute{1508424}{Laboratoire Charles Coulomb, CNRS and University of Montpellier, France};
\iinstitute{903287}{Physics Department, University of Trieste, Italy};
\iinstitute{903889}{Kavli Institute for Theoretical Physics, University of California, Santa Barbara, United States};
\iinstitute{903237}{SUNY at Stony Brook, United States};
\iinstitute{902999}{School of Physics, University of Melbourne, Australia};
\iinstitute{903299}{UC Berkeley, United States};
\iinstitute{902893}{Iowa State University, United States};
\iinstitute{902865}{Illinois Institute of Technology, United States};
\iinstitute{902767}{Delhi University, India};
\iinstitute{903940}{{College of Engineering, Design and Physical Sciences, Brunel University, United Kingdom}};
\iinstitute{1424710}{Center for Quantum Mathematics and Physics (QMAP), University of California, Davis, United States};
\iinstitute{903142}{Purdue University, United States};
\iinstitute{907933}{Theoretische Natuurkunde and IIHE/ELEM, Vrije Universiteit Brussel, Belgium};
\iinstitute{903009}{{Dipartimento di Fisica Aldo Pontremoli, Universit\'a degli Studi di Milano, Italy}};
\iinstitute{902804}{Physics Department, University of Florida, United States};
\iinstitute{905190}{Universitat de Barcelona, Spain};
\iinstitute{902711}{California Institute of Technology, United States};
\iinstitute{16750}{Princeton University, United States};
\iinstitute{903302}{UC, Irvine, United States};
\iinstitute{902878}{INFN Sezione di Bologna, Italy};
\iinstitute{902748}{{Department of Physics, University of Colorado, United States}};
\iinstitute{903326}{University of Vienna, Faculty of Physics, Austria};
\iinstitute{904961}{JLab, United States};
\iinstitute{902813}{D\'epartment de Physique Th\'eorique, Universit\'e de Gen\`eve, Switzerland};
\iinstitute{1189711}{Lawrence Berkeley National Laboratory, United States};
\iinstitute{411233}{International Institute of Physics, Universidade Federal do Rio Grande do Norte, Brazil};
\iinstitute{909079}{{CAFPE}, Spain};
\iinstitute{903324}{Institut f\"ur Hochenergiephysik der \"Osterreichischen Akademie der Wissenschaften, Austria};
\iinstitute{903006}{Michigan State University, United States};
\iinstitute{902823}{School of Physics and Astronomy, University of Glasgow, United Kingdom};
\iinstitute{1273509}{University of Oklahoma, United States};
\iinstitute{902883}{Universit{\'a} di Napoli ``Federico II" and INFN Napoli, Italy};
\iinstitute{902984}{University of Manchester, United Kingdom};
\iinstitute{904214}{Physics, SUPA, United Kingdom};
\iinstitute{902916}{{High Energy Accelerator Research Organization KEK, Japan}};
\iinstitute{903268}{Tohoku University, Japan};
\iinstitute{903092}{Ohio State University, United States};
\iinstitute{903119}{LPNHE, Sorbonne Universit\'e, France};
\iinstitute{905405}{Paul Scherrer Institute, Switzerland};
\iinstitute{903170}{{Royal Holloway University of London, Department of Physics, United Kingdom}};
\iinstitute{911212}{Institut de Ciencies del Cosmos (ICC), Spain};
\iinstitute{904536}{Atominstitut, Technische Universit\"at Wien, Austria}
}

\begin{titlepage}

\vspace*{-1.8cm}

\noindent
\begin{tabular*}{\linewidth}{lc@{\extracolsep{\fill}}r@{\extracolsep{0pt}}}
\vspace*{-1.2cm}\mbox{\!\!\!\includegraphics[width=.14\textwidth]{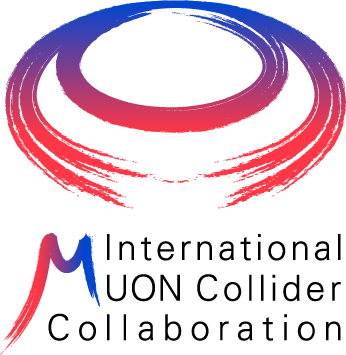}} & &  \\
 & & \today \\  
 & & \href{https://muoncollider.web.cern.ch}{https://muoncollider.web.cern.ch} \\ 
 & & \\
\hline
\end{tabular*}

\vspace*{0.3cm}
%
%
\maketitle
\vspace{\fill}







\end{titlepage}

\setcounter{tocdepth}{3}

\section{Overview}\label{intro}

Colliders are microscopes that explore the structure and the interactions of particles at the shortest possible length scale. Their goal is not to chase discoveries that are inevitable or perceived as such based on current knowledge. On the contrary, their mission is to explore the unknown in order to acquire radically novel knowledge.

The current experimental and theoretical situation of particle physics is particularly favorable to collider exploration. No inevitable discovery diverts our attention from pure exploration, and we can focus on the basic questions that best illustrate our ignorance. Why is electroweak symmetry broken and what sets the scale? Is it really broken by the Standard Model Higgs or by a more rich Higgs sector? Is the Higgs an elementary or a composite particle? Is the top quark, in light of its large Yukawa coupling, a portal towards the explanation of the observed pattern of flavor? Is the Higgs or the electroweak sector connected with dark matter? Is it connected with the origin of the asymmetry between baryons and anti-baryons in the Universe?

The next collider should offer broad and varied opportunities for exploration. It should deepen our understanding of the questions above, and be ready to tackle novel challenges that might emerge from future discoveries at the LHC or other experiments. The current $g$-2 and lepton flavor universality violation anomalies, which are both related to muons, are examples of tensions with the Standard Model (SM) that the next collider might be called to elucidate, by accessing the corresponding microscopic explanation.

A comprehensive exploration must exploit the complementarity between energy and precision. Precise measurements allow us to study the dynamics of the particles we already know, looking for the indirect manifestation of yet unknown new physics. With a very high energy collider we can access the new physics particles directly. These two exploration strategies are normally associated with two distinct machines, either colliding electrons/positrons ($ee$) or protons ($pp$). 

With muons instead, both strategies can be effectively pursued at a single collider that combines the advantages of $ee$ and of $pp$ machines. Moreover the simultaneous availability of energy and precision offers unique perspectives of indirect sensitivity to new physics at the $100$~TeV scale, as well as unique perspectives for the characterization of new heavy particles discovered at the muon collider itself. This is the picture that emerges from the studies of the muon colliders physics potential performed so far, to be reviewed in this document.

\section{Why muons?}\label{whym}

Muons, like protons, can be made to collide with a center of mass energy of $10$~TeV or more in a relatively compact ring, without fundamental limitations from synchrotron radiation. However, being point-like particles, unlike protons, their nominal center of mass collision energy $E_{\rm{cm}}$ is entirely available to produce high-energy reactions that probe lengths scale as short as $1/E_{\rm{cm}}$. The relevant energy for proton colliders is instead the center of mass energy of the collisions between the partons that constitute the protons. The partonic collision energy is distributed statistically, and approaches a significant fraction of the proton collider nominal energy with very low probability. A muon collider with a given nominal energy and luminosity is thus evidently way more effective than a proton collider with comparable energy and luminosity.

\begin{figure}
\begin{minipage}{0.48\textwidth}
\begin{center}
\includegraphics[width=\textwidth]{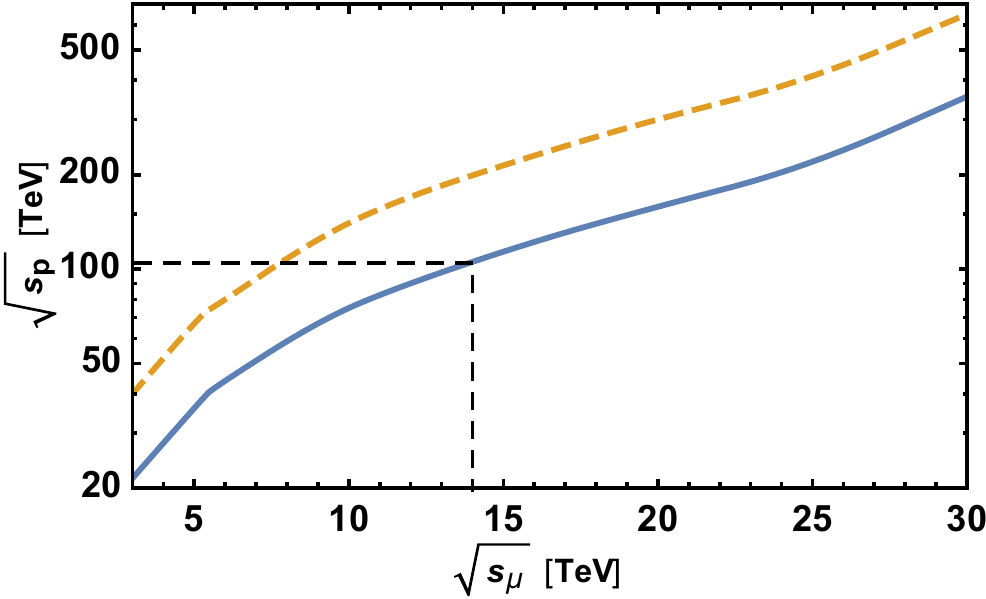}
\end{center}
\end{minipage}
\hfill
\begin{minipage}{0.47\textwidth}
\begin{center}
\includegraphics[width=\textwidth]{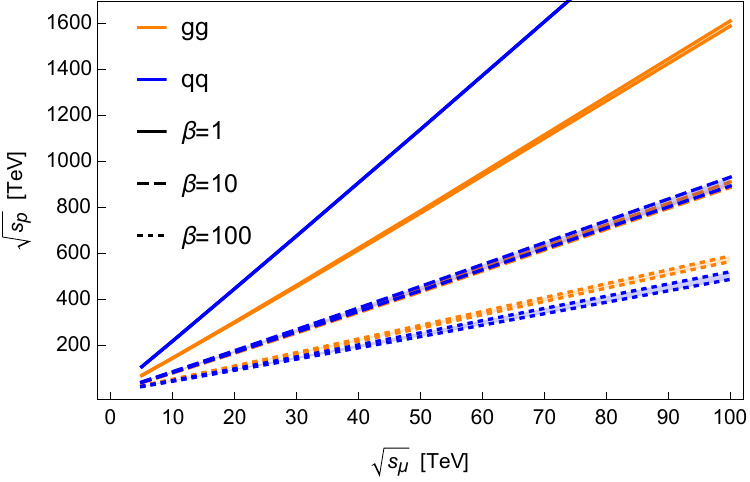}
\end{center}
\end{minipage}
\caption{Equivalent proton collider energy. The left plot~\cite{Delahaye:2019omf}, assumes that $qq$ and~$gg$~partonic initial states both contribute to the production. In the orange and blue lines, $\beta=1$ and $\beta=10$, respectively. In the right panel~\cite{AlAli:2021let}, production from $qq$ and from $gg$ are considered separately.
\label{pvsm}}
\end{figure}

This concept is made quantitative in Figure~\ref{pvsm}. The figure displays the center of mass energy \parbox[c][0pt]{19pt}{${\sqrt{s\,}}_{\hspace{-2pt}p}$} that a proton collider must possess to be ``equivalent'' to a muon collider of a given energy \parbox[c][0pt]{54pt}{$E_{\rm{cm}}=\sqrt{s\,}_{\hspace{-2pt}\mu}$}. Equivalence is defined~\cite{Delahaye:2019omf,Costantini:2020stv,AlAli:2021let} in terms of the pair production cross-section for heavy particles, with mass close to the muon collider kinematical threshold of \parbox[c][0pt]{32pt}{$\sqrt{s\,}_{\hspace{-2pt}\mu}/2$}. The equivalent \parbox[c][0pt]{19pt}{${\sqrt{s\,}}_{\hspace{-2pt}p}$} is the proton collider center of mass energy for which the cross-sections at the two colliders are equal. 

The estimate of the equivalent \parbox[c][0pt]{19pt}{${\sqrt{s\,}}_{\hspace{-2pt}p}$} depends on the relative strength $\beta$ of the heavy particle interaction with the partons and with the muons. If the heavy particle only possesses electroweak quantum numbers, $\beta=1$ is a reasonable estimate because the particles are produced by the same interaction at the two colliders. If instead it also carries QCD color, the proton collider can exploit the QCD interaction to produce the particle, and a ratio of $\beta=10$ should be considered owing to the large QCD coupling and color factors. The orange line on the left panel of Figure~\ref{pvsm}, obtained with $\beta=1$, is thus representative of purely electroweak particles. The blue line, with $\beta=10$, is instead a valid estimate for particles that also possess QCD interactions, as it can be verified in concrete examples.

The general lesson we learn from the left panel of Figure~\ref{pvsm} (orange line) is that at a proton collider with around $100$~TeV energy the cross-section for processes with an energy threshold of around $10$~TeV is much smaller than the one of a muon collider operating at  \parbox[c][0pt]{54pt}{$E_{\rm{cm}}={\sqrt{s\,}}_{\hspace{-2pt}\mu}$} $\sim10$~TeV. The gap can be compensated only if the process dynamics is different and more favorable at the proton collider, like in the case of QCD production. The general lesson has been illustrated for new heavy particles production, where the threshold is provided by the particle mass. But it also holds for the production of light SM particles with energies as high as $E_{\rm{cm}}$, which are very sensitive indirect probes of new physics. This makes exploration by high energy measurements more effective at muon than at proton colliders, as we will see in Section~\ref{HEM}. Moreover the large luminosity for high energy muon collisions produces the copious emission of effective vector bosons. In turn, they are responsible at once for the tremendous direct sensitivity of muon colliders to ``Higgs portal'' type new physics and for their excellent perspectives to measure single and double Higgs couplings precisely as we will see in Section~\ref{dirr} and~\ref{VBF}, respectively.

On the other hand, no quantitative conclusion can be drawn from Figure~\ref{pvsm} on the comparison between the muon and proton colliders discovery reach for the heavy particles. That assessment will be performed in the following section based on available proton colliders projections.

\section{Direct reach}\label{dirr}

\begin{figure}
\begin{minipage}{0.52\textwidth}
\begin{center}
\includegraphics[width=\textwidth]{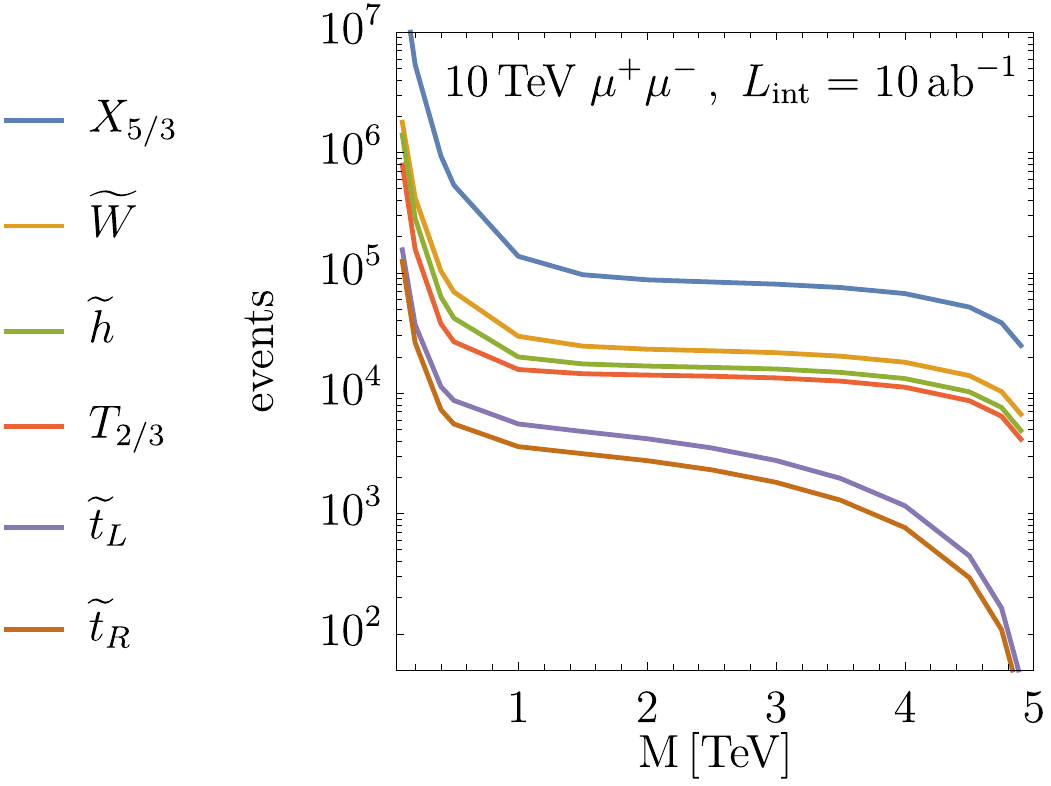}
\end{center}
\end{minipage}
\hfill
\begin{minipage}{0.38\textwidth}
\begin{center}
\includegraphics[width=\textwidth]{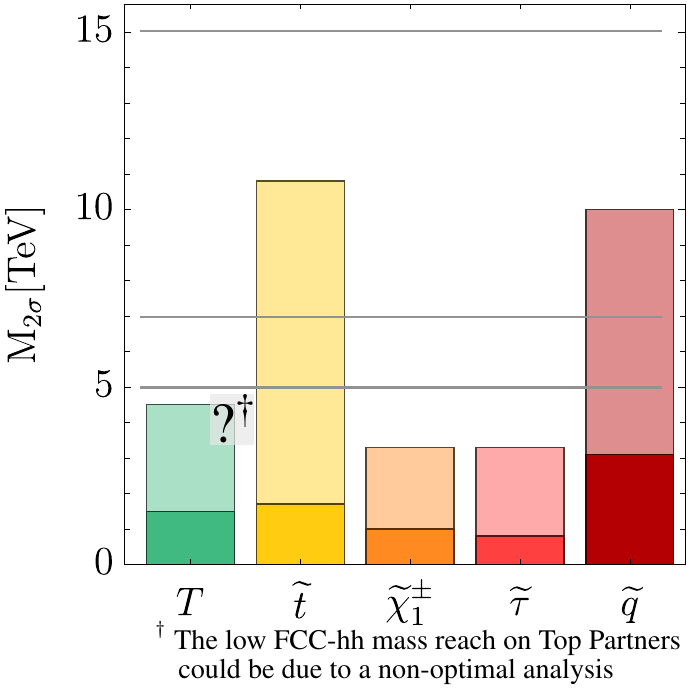}
\end{center}
\end{minipage}
\caption{Left panel: the number of expected events (from Ref.~\cite{Buttazzo:2020uzc}, see also~\cite{Costantini:2020stv}) at a $10$~TeV muon collider, with  $10$~ab$^{-1}$ luminosity, for several BSM particles. Right panel: $95\%$~CL mass reach, from Ref.~\cite{EuropeanStrategyforParticlePhysicsPreparatoryGroup:2019qin}, at the HL-LHC (solid bars) and at the FCC-hh (shaded bars). The tentative discovery reach of a 10, 14 and 30~TeV muon collider are reported as horizontal lines.
\label{dr}}
\end{figure}

The left panel of Figure~\ref{dr} displays the number of expected events, at a $10$~TeV muon collider with $10$~ab$^{-1}$ integrated luminosity, for the pair production due to electroweak interactions of Beyond the Standard Model (BSM) particles with variable mass M. The particles are named with a standard BSM terminology, however the results do not depend on the detailed BSM model (such as Supersymmetry or Composite Higgs) in which these particles emerge, but only on their Lorentz and gauge quantum numbers. The dominant production mechanism at high mass is the direct $\mu^+\mu^-$ annihilation, whose cross-section flattens out below the kinematical threshold at ${\rm{M}}=5$~TeV. The cross-section increase at low mass is due to the production from effective vector bosons annihilation.

The figure shows that with the target luminosity of $10$~ab$^{-1}$ a $E_{\rm{cm}}=10$~TeV muon collider can produce the BSM particles abundantly. If they decay to energetic and detectable SM final states, the new particles can be definitely discovered up to the kinematical threshold. Taking into account that entire target integrated luminosity will be collected in $5$ years, a few months of run could be sufficient for a discovery. Afterwards, the large production rate will allow us to observe the new particles decaying in multiple final states and to measure kinematical distributions. We will thus be in the position of characterizing the properties of the newly discovered states precisely. Similar considerations hold for muon colliders with higher $E_{\rm{cm}}$, up to the fact that the kinematical mass threshold obviously grows to $E_{\rm{cm}}/2$. Notice however that the production cross-section decreases as $1/E_{\rm{cm}}^2$.\footnote{The scaling is violated by the vector boson annihilation channel, which however is relevant only at low mass.} 
Therefore we obtain as many events as in the left panel of Figure~\ref{dr} only if the integrated luminosity grows as
\begin{equation}\label{lums}
\displaystyle
L_{\rm{int}}=10\,{\rm{ab}}^{-1}\left(\frac{E_{\rm{cm}}}{10\,{\rm{TeV}}}\right)^2\,.
\end{equation}
A luminosity that is lower than this by a factor of around $10$ would not affect the discovery reach, but it might, in some cases, slightly reduce the potential for characterizing the discoveries.

The direct reach of muon colliders vastly and generically exceeds the sensitivity of the High-Luminosity LHC (HL-LHC). This is illustrated by the solid bars on the right panel of Figure~\ref{dr}, where we report the projected HL-LHC mass reach~\cite{EuropeanStrategyforParticlePhysicsPreparatoryGroup:2019qin} on several BSM states. The $95\%$~CL exclusion is reported, instead of the discovery, as a quantification of the physics reach. Specifically, we consider Composite Higgs fermionic top-partners $T$ (e.g., the \parbox[c][0pt]{21pt}{$X_{5/3}$} and the \parbox[c][0pt]{19pt}{$T_{2/3}$}) and supersymmetric particles such as stops~\raisebox{2pt}{\parbox[c][0pt]{6pt}{${\widetilde{t}}$}}, charginos \raisebox{2pt}{\parbox[c][0pt]{14pt}{${\widetilde{\chi}}_1^\pm$}}, stau leptons~\raisebox{2pt}{\parbox[c][0pt]{6pt}{${\widetilde{\tau}}$}} and squarks~\raisebox{1pt}{\parbox[c][0pt]{6pt}{${\widetilde{q}}$}}. For each particle we report the highest possible mass reach, as obtained in the configuration for the BSM particle couplings and decay chains that maximizes the hadron colliders sensitivity. The reach of a $100$~TeV proton-proton collider (FCC-hh) is shown as shaded bars on the same plot. The muon collider reach, displayed as horizontal lines for $E_{\rm{cm}}=10$, $14$ and~$30$~TeV, exceeds the one of the FCC-hh for several BSM candidates and in particular, as expected, for purely electroweak charged states.

\begin{figure}
\begin{minipage}{0.5\textwidth}
\begin{center}
\includegraphics[width=1\textwidth]{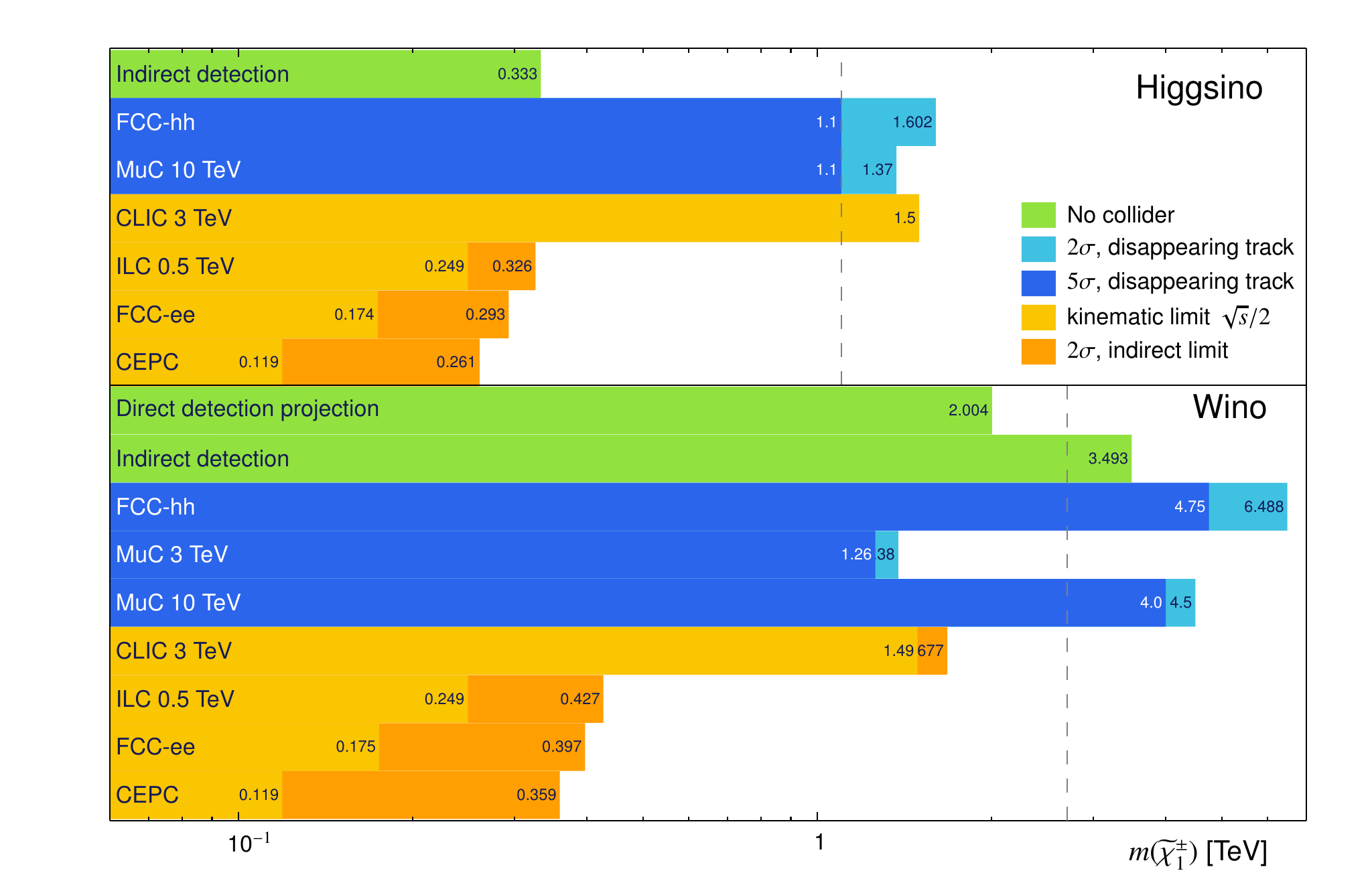}
\vspace{-4pt}
\end{center}
\end{minipage}
\hfill
\begin{minipage}{0.51\textwidth}
\begin{center}
\includegraphics[width=\textwidth]{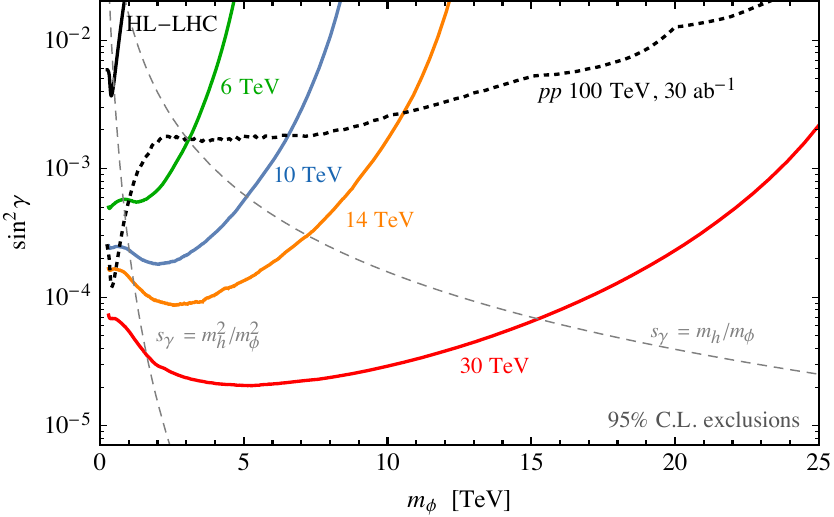}
\vspace{-18pt}
\end{center}
\end{minipage}
\caption{Left panel: exclusion and discovery mass reach on Higgsino and Wino Dark Matter candidates at muon colliders from disappearing tracks, and at other facilities. The plot is adapted from Ref.~\cite{Capdevilla:2021fmj}. Right: exclusion contour~\cite{AlAli:2021let} for a scalar singlet of mass $m_\phi$ mixed with the Higgs boson with strength $\sin\gamma$
\label{dm}}
\end{figure}

Several interesting BSM particles do not decay to easily detectable final states, and an assessment of their observability requires dedicated studies. A clear case is the one of minimal WIMP Dark Matter (DM) candidates (see e.g.~\cite{AlAli:2021let} and references therein). The charged state in the DM electroweak multiplet is copiously produced, but it decays to the invisible DM plus a soft undetectable pion, owing to the small mass-splitting. WIMP DM can be studied at muon colliders in several channels (such as mono-photon) without directly observing the charged state~\cite{Han:2020uak,Bottaro:2021snn}. Alternatively, one can instead exploit the disappearing tracks produced by the charged particle~\cite{Capdevilla:2021fmj}. The result is displayed on the left panel of Figure~\ref{dm} for the simplest candidates, known as Higgsino and Wino. A $10$~TeV muon collider reaches the ``thermal'' mass, marked with a dashed line, for which the observed relic abundance is obtained by thermal freeze out. Other minimal WIMP candidates become kinematically accessible at higher muon collider energies~\cite{Han:2020uak,Bottaro:2021snn}. Muon colliders could actually even probe some of these candidates when they are above the kinematical threshold, by studying their indirect effects on high-energy SM processes~\cite{DiLuzio:2018jwd,RFXZ}.

New physics particles are not necessarily coupled to the SM by gauge interaction. One setup that is relevant in several BSM scenarios (including models of baryogenesis, dark matter, and neutral naturalness) is the ``Higgs portal'' one, where the BSM particles interact most strongly with the Higgs field. By the Goldstone Boson Equivalence Theorem, Higgs field couplings are interactions with the longitudinal polarizations of the SM massive vector bosons $W$ and $Z$, which enable Vector Boson Fusion (VBF) production of the new particles. A muon collider is extraordinarily sensitive to VBF production, owing to the large luminosity for effective vector bosons. This is illustrated on the right panel of Figure~\ref{dm}, in the context of a benchmark model~\cite{Buttazzo:2018qqp,AlAli:2021let} (see also \cite{Ruhdorfer:2019utl,Liu:2021jyc}) where the only new particle is a real scalar singlet with Higgs portal coupling. The coupling strength is traded for the strength of the mixing with the Higgs particle, $\sin\gamma$, that the interaction induces. The scalar singlet is the simplest extension of the Higgs sector. Extensions with richer structure, such as involving a second Higgs doublet, are a priori easier to detect as one can exploit the electroweak production of the new charged Higgs bosons, as well as their VBF production. See Ref.s~\cite{Han:2021udl,Chakrabarty:2014pja,Kalinowski:2020rmb} for dedicated studies, and Ref.~\cite{MuonCollider:2022xlm} for a review.

We have seen that in several cases the muon collider direct reach compares favorably to the one of the most ambitious future proton collider project. This is not a universal statement, in particular it is obvious that at a muon collider it is difficult to access heavy particles that carry only QCD interactions. One might also expect a muon collider of $10$~TeV to be generically less effective than a $100$~TeV proton collider for the detection of particles that can be produced singly. For instance, for additional $Z'$ massive vector bosons, that can be probed at the FCC-hh well above the $10$~TeV mass scale. We will see in Section~\ref{HEM} that the situation is slightly more complex and that, in the case of $Z'$s, a $10$~TeV muon collider sensitivity actually exceeds the one of the FCC-hh dramatically (see the right panel of Fig.~\ref{fig:HEM}).

\section{A vector bosons collider}\label{VBF}

When two electroweak charged particles like muons collide at an energy much above the electroweak scale $m_{{\rm{\textsc{w}}}}\sim100~$GeV, they have a high probability to emit ElectroWeak (EW) radiation. There are multiple types of EW radiation effects that can be observed at a muon collider and play a major role in muon collider physics. Actually we will argue in Section~\ref{EWRadiation} that the experimental observation and the theoretical description of these phenomena emerges as a self-standing reason of scientific interest in muon colliders. 

\begin{figure}
\centering{
\begin{minipage}{0.35\textwidth}
\vspace{-10pt}
\includegraphics[width=\textwidth]{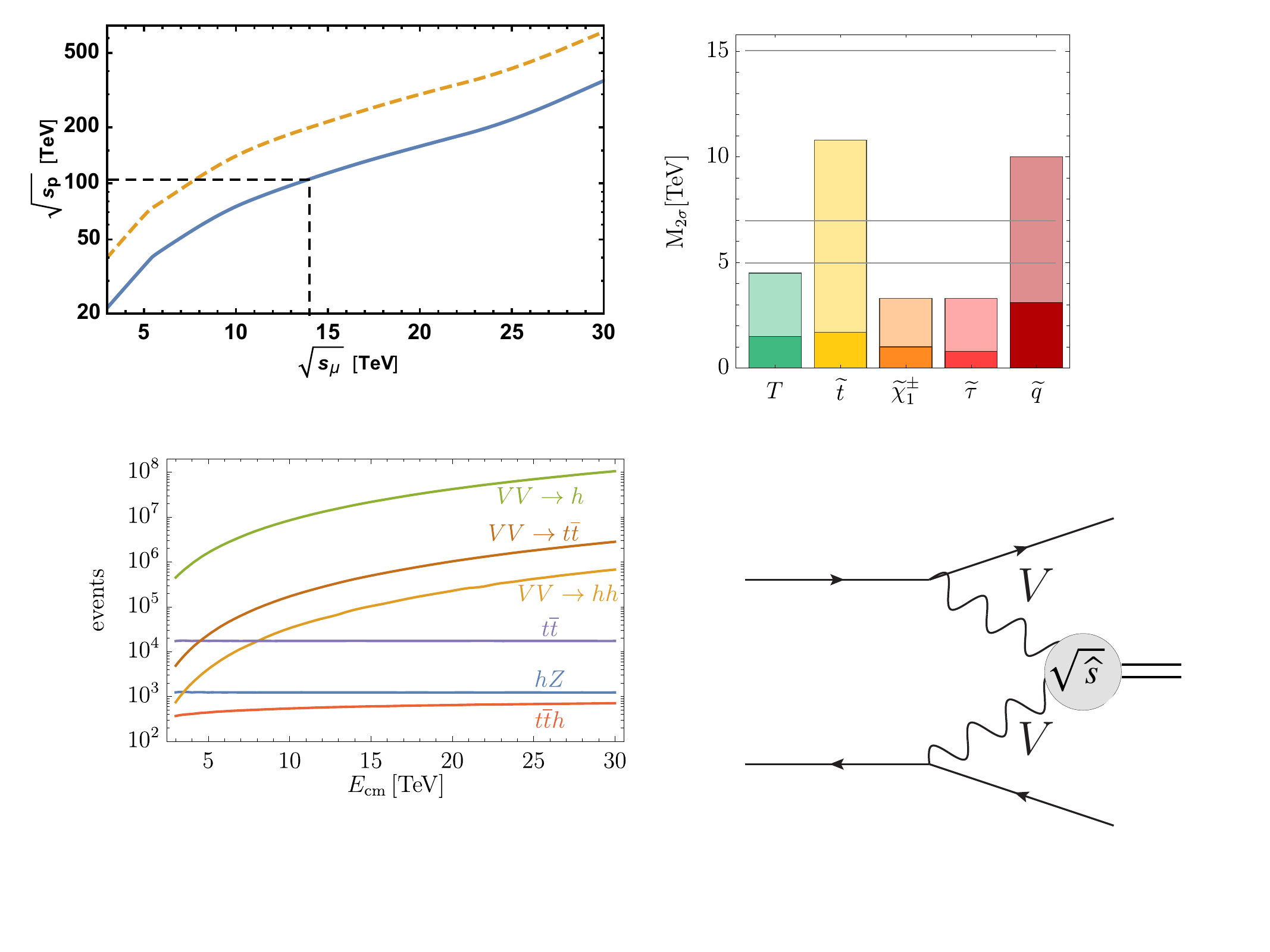}
\end{minipage}
\hspace{30pt}
\begin{minipage}{0.5\textwidth}
\vspace{0pt}
\includegraphics[width=0.95\textwidth]{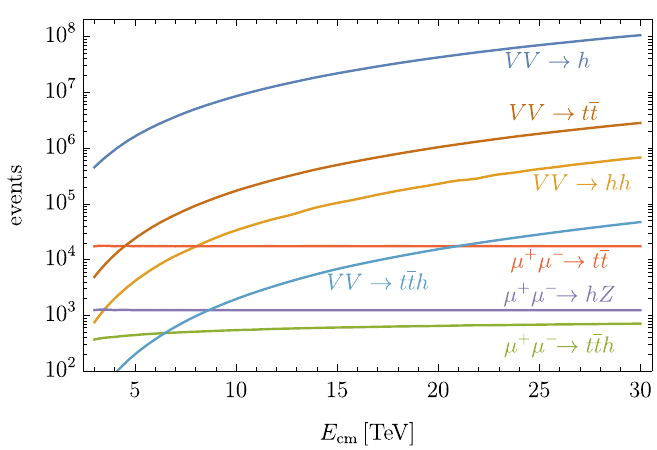}
\end{minipage}}
\caption{Left panel: schematic representation of vector boson fusion or scattering processes. The collinear $V$ bosons emitted from the muons participate to a process with hardness \mbox{$\sqrt{\hat{s}}\ll E_{\rm{cm}}$}. Right panel: number of expected events for selected SM processes at a muon collider with variable $E_{\rm{cm}}$ and luminosity scaling as in eq.~(\ref{lums}). 
\label{fig:EV}}
\end{figure}

Here we focus on the practical implications~\cite{Delahaye:2019omf,Costantini:2020stv,Han:2020pif,Buttazzo:2020uzc,AlAli:2021let,Forslund:2022_mu3_10} of the collinear emission of nearly on-shell massive vector bosons, which is the analog in the EW context of the Weizsaecker--Williams emission of photons. The vector bosons $V$ participate, as depicted in Figure~\ref{fig:EV}, to a scattering process with a hard scale \parbox[c][0pt]{15pt}{$\sqrt{\hat{s}}$} that is much lower than the muon collision energy \parbox[c][0pt]{20pt}{$E_{\rm{cm}}$}. The typical cross-section for $VV$ annihilation processes is thus enhanced by \parbox[c][0pt]{30pt}{$E_{\rm{cm}}^2/{\hat{s}}$}, relative to the typical cross-section for $\mu^+\mu^-$ annihilation, whose hard scale is instead $E_{\rm{cm}}$. The emission of the $V$ bosons from the muons is suppressed by the EW coupling, but the suppression is mitigated or compensated by logarithms of the separation between the EW scale and $E_{\rm{cm}}$ (see~\cite{Costantini:2020stv,AlAli:2021let} for a pedagogical overview). The net result is a very large cross-section for VBF processes that occur at \parbox[c][0pt]{46pt}{$\sqrt{\hat{s}}\sim m_{\rm{\textsc{w}}}$}, with a tail in \parbox[c][0pt]{17pt}{$\sqrt{\hat{s}}$} up to the TeV scale.

We already emphasized (see Figure~\ref{dr}) the importance of VBF for the direct production of new physics particles. The relevance of VBF for probing new physics indirectly simply stems for the huge rate of VBF SM processes, summarized on the right panel of Figure~\ref{fig:EV}. In particular we see that a $10$~TeV muon collider produces ten million Higgs bosons, which is around $10$ times more than future $e^+e^-$ Higgs factories. Since the Higgs bosons are produced in a relatively clean environment, a $10$~TeV muon collider (over-)qualifies as a Higgs factory~\cite{Forslund:2022_mu3_10,AlAli:2021let,Han:2020pif,Bartosik:2019dzq,Bartosik:2020xwr}. Unlike $e^+e^-$ Higgs factories, a muon collider also produces Higgs pairs copiously, enabling accurate measurements of the Higgs trilinear coupling~\cite{Costantini:2020stv,Han:2020pif,Buttazzo:2020uzc} and possibly also of the quadrilinear coupling~\cite{Chiesa:2020awd}. 

\begin{figure}
\begin{minipage}{0.55\textwidth}
\renewcommand{\arraystretch}{.89}
\setlength{\arrayrulewidth}{.2mm}
\setlength{\tabcolsep}{0.6 em}
\begin{center}
\begin{tabular}{c|c|c|c}
\hline
& \multicolumn{1}{c|}{\makebox[35pt]{\small{HL-LHC}}} & \makebox[35pt]{\small{HL-LHC}} & \makebox[35pt]{\small{HL-LHC}} \\[-2pt]
& \multicolumn{1}{c|}{\ } & \multicolumn{1}{l|}{\makebox[35pt]{\small{$+10\,\textrm{{TeV}}$}}} & \multicolumn{1}{l}{\makebox[35pt]{\small{$+10\,\textrm{{TeV}}$}}}  \\[-4pt]
\ & \ & \multicolumn{1}{c|}{\ } & \multicolumn{1}{l}{\hspace{-2pt}$+\,e e$} \\ 
\hline
$\kappa_W$ & 1.7 & 0.1 & 0.1 \\ \hline
$\kappa_Z$ & 1.5 & 0.4 & 0.1 \\ \hline
$\kappa_g$ & 2.3 & 0.7 & 0.6 \\ \hline
$\kappa_{\gamma}$ & 1.9 & 0.8 & 0.8 \\ \hline
$\kappa_{Z\gamma}$ & 10 & 7.2 & 7.1 \\ \hline 
$\kappa_c$ & - & 2.3 & 1.1 \\ \hline
$\kappa_b$ & 3.6 & 0.4 & 0.4\\ \hline
$\kappa_{\mu}$ & 4.6 & 3.4 & 3.2 \\ \hline
$\kappa_{\tau}$ & 1.9 & 0.6 & 0.4 \\ \hline\hline
$\kappa_t^*$ & 3.3 & 3.1 & 3.1 \\ \hline
\multicolumn{4}{l}{ {\scriptsize $^*$ No input used for $\mu$ collider}}\\
\end{tabular}
\end{center}
\end{minipage}
\hfill
\begin{minipage}{0.45\textwidth}
\vspace{0pt}
\includegraphics[width=.93\textwidth]{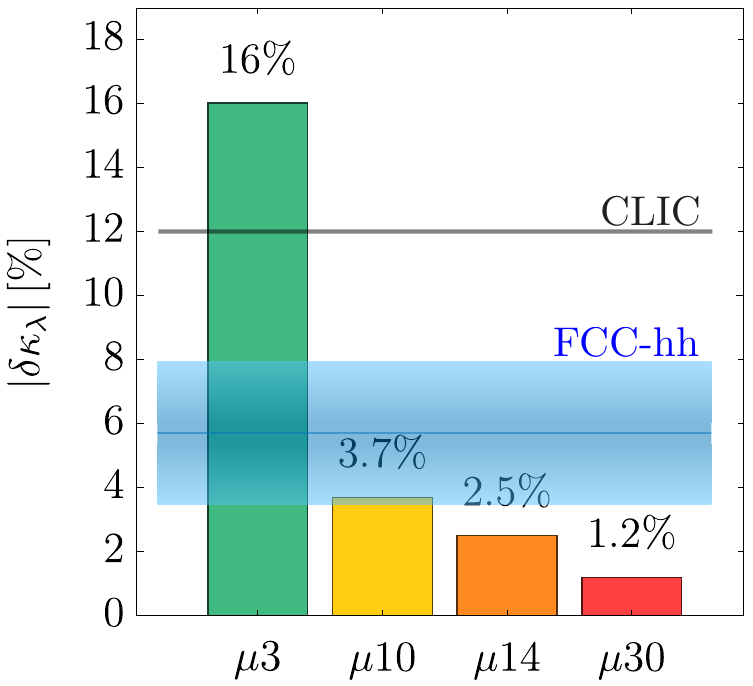}
\end{minipage}
\caption{Left panel: $1\sigma$ sensitivities (in \%) from a 10-parameter fit in the $\kappa$-framework at a $10$~TeV muon collider with $10$~ab$^{-1}$~\cite{MuonCollider:2022xlm}, compared with HL-LHC. The effect of measurements from a $250$~GeV $e^+e^-$ Higgs factory is also reported. Right panel: sensitivity to $\delta\kappa_\lambda$ for different $E_{\rm{cm}}$. The luminosity is as in eq.~(\ref{lums}) for all energies, apart from $E_{\rm{cm}}\hspace{-2pt}=\hspace{-2pt}3$~TeV, where doubled luminosity (of 1.8~ab$^{-1}$) is assumed~\cite{MuonCollider:2022xlm}.
\label{tab:higgscouplingfit}}
\end{figure}

The opportunities for Higgs physics at a muon collider are summarized elsewhere~\cite{MuonCollider:2022xlm}. In Figure~\ref{tab:higgscouplingfit} we report for illustration the results of a 10-parameter fit to the Higgs couplings in the $\kappa$-framework at a $10$~TeV muon collider, and the sensitivity projections on the anomalous Higgs trilinear coupling $\delta\kappa_\lambda$. The table shows that a $10$~TeV muon collider will improve significantly and broadly our knowledge of the properties of the Higgs boson. The combination with the measurements performed at an $e^+e^-$ Higgs factory, reported on the third column, does not affect the sensitivity to several couplings appreciably, showing the good precision that a muon collider alone can attain. However, it also shows complementarity with an $e^+e^-$ Higgs factory program. More examples of this complementarity are discussed in~\cite{MuonCollider:2022xlm}. 

In the right panel of the figure we see that the performances of muon colliders in the measurement of $\delta\kappa_\lambda$ are similar or much superior to the one of the other future colliders where this measurement could be performed. In particular, CLIC measures $\delta\kappa_\lambda$ at the $10\%$ level~\cite{deBlas:2018mhx}, and the FCC-hh sensitivity ranges from $3.5$ to $8\%$ depending on detector assumptions~\cite{Mangano:2020sao}. A determination of $\delta\kappa_\lambda$ that is way more accurate than the HL-LHC projections is possible already at a low energy stage of a muon collider with $E_{\rm{cm}}=3$~TeV.

The potential of a muon collider as a vector boson collider has not been explored fully. In particular a systematic investigation of vector boson scattering processes, such as $WW\hspace{-3pt}\to\hspace{-3pt} WW$, has not been performed. The key role played by the Higgs boson to eliminate the energy growth of the corresponding Feynman amplitudes could be directly verified at a muon collider by means of differential measurements that extend well above one TeV for the invariant mass  of the scattered vector bosons. Along similar lines, differential measurements of the $WW\hspace{-3pt}\to\hspace{-3pt} HH$ process has been studied in~\cite{Buttazzo:2020uzc,Han:2020pif} (see also~\cite{Costantini:2020stv}) as an effective probe of the composite nature of the Higgs boson, with a reach that is comparable or superior to the one of Higgs coupling measurements. A similar investigation was performed in~\cite{AlAli:2021let,Costantini:2020stv} (see also~\cite{Costantini:2020stv}) for $WW\hspace{-3pt}\to\hspace{-3pt} t{\overline{t}}$, aimed at probing Higgs-top interactions.

\section{High-energy measurements}\label{HEM}

Direct $\mu^+\mu^-$ annihilation, such as $HZ$ and $t{\overline{t}}$ production reported in Figure~\ref{fig:EV}, displays a number of expected events of the order of several thousands. These are much less than the events where a Higgs or a $t{\overline{t}}$ pair are produced from VBF, but they are sharply different and easily distinguishable. The invariant mass of the particles produced by direct annihilation is indeed sharply peaked at the collider energy $E_{\rm{cm}}$, while the invariant mass rarely exceeds one tenth of $E_{\rm{cm}}$ in the VBF production mode. 

The good statistics and the limited or absent background thus enables percent of few-percent level measurements of SM cross sections for hard scattering processes of energy $E_{\rm{cm}}=10$~TeV or more. An incomplete list of the many possible measurements is provided in Ref.~\cite{Chen:2022msz}, including the resummed effects of EW radiation on the cross section predictions. It is worth emphasizing that also charged final states such as $WH$ or $\ell\nu$ are copiously produced at a muon collider. The electric charge mismatch with the neutral $\mu^+\mu^-$ initial state is compensated by the emission of soft and collinear $W$ bosons, that occurs with high probability because of the large energy.

High energy scattering processes are as unique theoretically as they are experimentally~\cite{Delahaye:2019omf,Buttazzo:2020uzc,Chen:2022msz}. They give direct access to the interactions among SM particles with $10$~TeV energy, which in turn provide indirect sensitivity to new particles at the $100$~TeV scale of mass. In fact, the effects on high-energy cross sections of new physics at energy $\Lambda\gg E_{\rm{cm}}$ generically scale as $(E_{\rm{cm}}/\Lambda)^2$ relative to the SM. Percent-level measurements thus give access to $\Lambda\sim100$~TeV. This is an unprecedented reach for new physics theories endowed with a reasonable flavor structure. Notice in passing that high-energy measurements are also useful to investigate flavor non-universal phenomena, as we will see below, and in Section~\ref{muspec}.

This mechanism is not novel. Major progress in particle physics always came from raising the available collision energy, producing either direct or indirect discoveries. For instance, precisely because of the quadratic energy scaling outlined above, the inner structure of nucleons and a first determination of their radius could be achieved only when the transferred energy in electron scattering could reach a significant fraction of the ``new physics'' scale $\Lambda=\Lambda_{\rm{\textsc{qcd}}}=300$~MeV~\cite{nobel}.  

\begin{figure}
\begin{minipage}{0.48\textwidth}
\includegraphics[width=\textwidth]{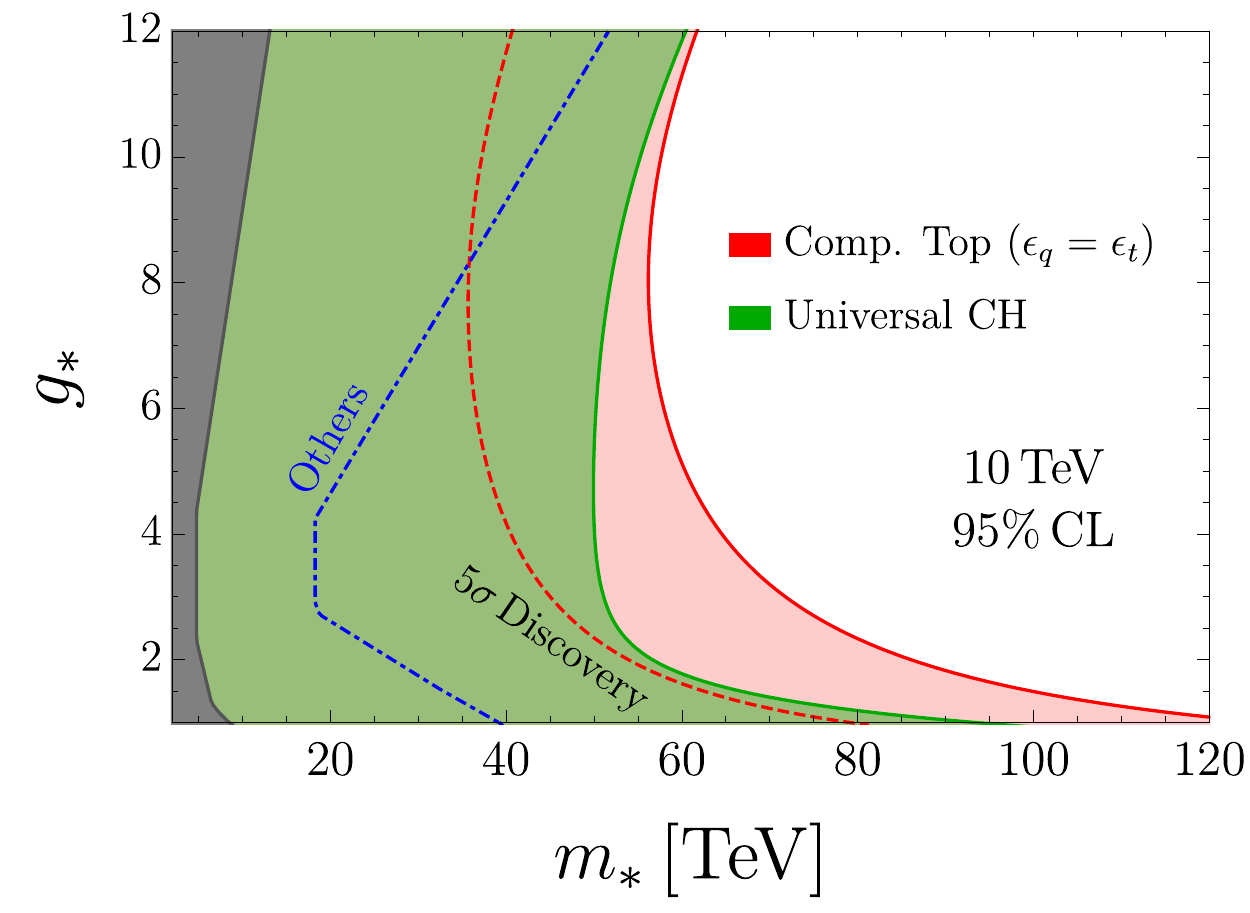}
\end{minipage}
\hfill
\begin{minipage}{0.48\textwidth}
\vspace{0pt}
\includegraphics[width=\textwidth]{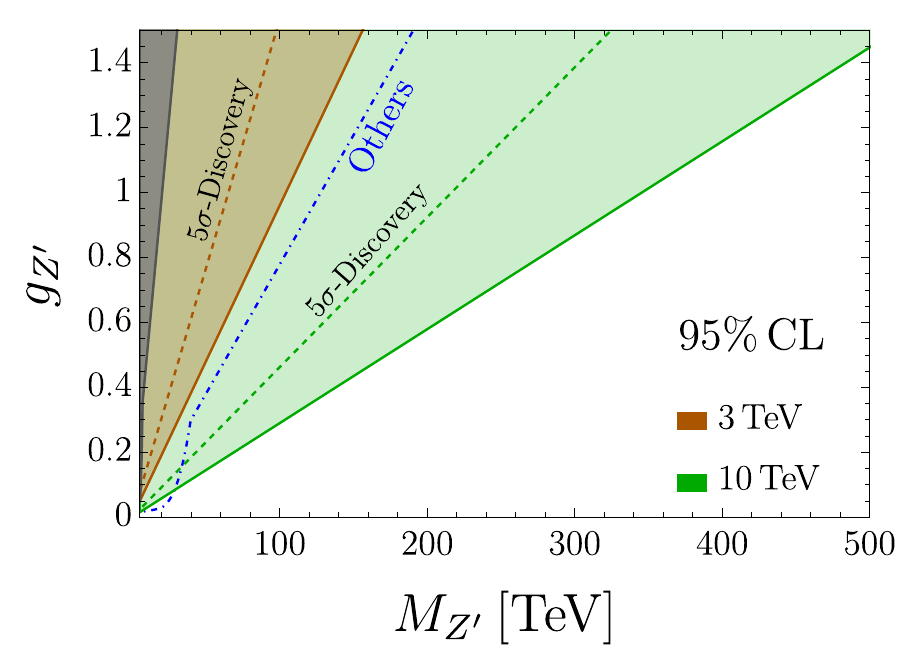}
\end{minipage}
\caption{Left panel: $95\%$ reach on the Composite Higgs scenario from high-energy measurements in di-boson and di-fermion final states~\cite{Chen:2022msz}. The green contour display the sensitivity from ``Universal'' effects related with the composite nature of the Higgs boson and not of the top quark. The red contour includes the effects of top compositeness. Right panel: sensitivity to a minimal $Z'$~\cite{Chen:2022msz}. Discovery contours at $5\sigma$ are also reported in both panels. 
\label{fig:HEM}}
\end{figure}

Figure~\ref{fig:HEM} illustrates the tremendous reach on new physics of a $10$~TeV muon collider with $10$~ab$^{-1}$ integrated luminosity. The left panel (green contour) is the sensitivity to a scenario that explains the microscopic origin of the Higgs particle and of the scale of EW symmetry breaking by the fact that the Higgs is a composite particle. In the same scenario the top quark is likely to be composite as well, which in turn explains its large mass and suggest a ``partial compositeness'' origin of the SM flavour structure. Top quark compositeness produces additional signatures that extend the muon collider sensitivity up to the red contour. The sensitivity is reported in the plane formed by the typical coupling $g_*$ and of the typical mass $m_*$ of the composite sector that delivers the Higgs. The scale $m_*$ physically corresponds to the inverse of the geometric size of the Higgs particle. The coupling $g_*$ is limited from around $1$ to $4\pi$, as in the figure. In the worst case scenario of intermediate $g_*$, a $10$~TeV muon collider can thus probe the Higgs radius up to the inverse of $50$~TeV, or discover that the Higgs is as tiny as $(35$~TeV$)^{-1}$. The sensitivity improves in proportion to the center of mass energy of the muon collider. 

The figure also reports, as blue dash-dotted lines denoted as ``Others'', the envelop of the $95\%$~CL sensitivity projections of all the future collider projects that have been considered for the~2020 update of the European Strategy for Particle Physics, summarized in Ref.~\cite{EuropeanStrategyforParticlePhysicsPreparatoryGroup:2019qin}. These lines include in particular the sensitivity of very accurate measurements at the EW scale performed at possible future $e^+e^-$ Higgs, Electroweak and Top factories. These measurements are not competitive because new physics at $\Lambda\sim100$~TeV produces unobservable one part per million effects on $100$~GeV energy processes. High-energy measurements at a $100$~TeV proton collider are also included in the dash-dotted lines. They are not competitive either, because the effective parton luminosity at high energy is much lower than the one of a $10$~TeV muon collider, as explained in Section~\ref{intro}. For example the cross-section for the production of an $e^+e^-$ pair with more than $9$~TeV invariant mass at the FCC-hh is of only $40$~ab, while it is of $900$~ab at a $10$~TeV muon collider. Even with a somewhat higher integrated luminosity, the FCC-hh just does not have enough statistics to compete with a $10$~TeV muon collider.

The right panel of Figure~\ref{fig:HEM} considers a simpler new physics scenario, where the only BSM state is a heavy $Z'$ spin-one particle. The ``Others'' line also includes the sensitivity of the FCC-hh from direct $Z'$ production. The line exceeds the $10$~TeV muon collider sensitivity contour (in green) only in a tiny region with $M_{Z'}$ around $20$~TeV and small $Z'$ coupling. This result substantiates our claim in Section~\ref{dirr} that a reach comparison based on the $2\hspace{-6pt}\to\hspace{-6pt}1$ single production of the new states is simplistic. Single $2\hspace{-3pt}\to\hspace{-4pt}1$ production couplings can produce indirect effect in $2\hspace{-3pt}\to\hspace{-3pt}2$ scattering by the virtual exchange of the new particle, and the muon collider is extraordinarily sensitive to these effects. Which collider wins is model-dependent. In the simple benchmark $Z'$ scenario, and in the motivated framework of Higgs compositeness that future colliders are urged to explore, the muon collider is just a superior device.

We have seen that high energy measurements at a muon collider enable the indirect discovery of new physics at a scale in the ballpark of $100$~TeV. However the muon collider also offers amazing opportunities for direct discoveries at a mass of several TeV, and unique opportunities to characterize the properties of the discovered particles, as emphasized in Section~\ref{dirr}. High energy measurements will enable us take one step further in the discovery characterization, by probing the interactions of the new particles well above their mass. For instance in the Composite Higgs scenario one could first discover Top Partner particles of few TeV mass, and next study their dynamics and their indirect effects on SM processes. This might be sufficient to pin down the detailed theoretical description of the newly discovered sector, which would thus be both discovered and theoretically characterized at the same collider. Higgs coupling determinations and other precise measurements that exploit the enormous luminosity for vector boson collisions, described in Section~\ref{VBF}, will also play a major role in this endeavour. 

Obviously, we can dream of such glorious outcome of the project only because energy and precision are simultaneously available at a muon collider.

\section{Muon-specific opportunities}\label{muspec}

In the quest for generic exploration, engineering collisions between muons and anti-muons for the first time is in itself a unique opportunity offered by the muon collider project. The concept can be made concrete by considering scenarios where the sensitivity to new physics stems from colliding muons, rather than electrons or other particles. An extensive overview of such ``muon-specific'' opportunities is provided in Ref.~\cite{MuonCollider:2022xlm}, based on the available literature~\cite{AlAli:2021let,Chakrabarty:2014pja,Capdevilla:2020qel, Buttazzo:2020ibd,Yin:2020afe, Capdevilla:2021rwo, Dermisek:2021ajd, Dermisek:2021mhi, Capdevilla:2021kcf,Huang:2021biu,Asadi:2021gah,Azatov:123,Huang:2021nkl, Homiller:123,Casarsa:2021rud,Han:2021lnp,muPDF,RKMC,Liu:2021akf,Cesarotti:2022ttv}. A concise summary is reported below.

It is perhaps worth emphasizing in this context that lepton flavour universality is not a fundamental property of Nature. Therefore new physics could exist, coupled to muons, that we could not yet discover using electrons. In fact, it is not only conceivable, but even expected that new physics could couple more strongly to muons than to electrons. Even in the SM lepton flavour universality is violated maximally by the Yukawa interaction with the Higgs field, that is larger for muons than for electrons. New physics associated to the Higgs or to flavour will most likely follow the same pattern, offering a competitive advantage of muon over electron collisions at similar energies. The comparison with proton colliders is less straightforward. By the same type of considerations one expects larger couplings with quarks, especially with the ones of the second and third generation. This expectation should be folded in with the much lower luminosity for heavier quarks at proton colliders than for muons at a muon collider. The perspectives of muon versus proton colliders are model-dependent and of course strongly dependent on the energy of the muon and of the proton collider.

The current $g$-2 and $B$-physics anomalies offer experimental hints for flavour non-universal new physics that point strongly and specifically to muons. The discrepancy of the muon $g$-2 measurements with the theoretical prediction is subject to intense investigation. If confirmed by further measurements and theoretical calculations, elucidating its origin might become a priority of particles physics in a few years' time. Similar considerations hold for the persistent flavour anomalies, including the most recent LHCb measurements of the $B$-meson decay ratios to muons over electrons, $R_{K^{(*)}}$. These anomalies will be further probed and potentially strengthened by the LHCb and Belle~II experiments on a timescale of few years.

\begin{figure}
\begin{minipage}{0.53\textwidth}
\begin{center}
\includegraphics[width=\textwidth]{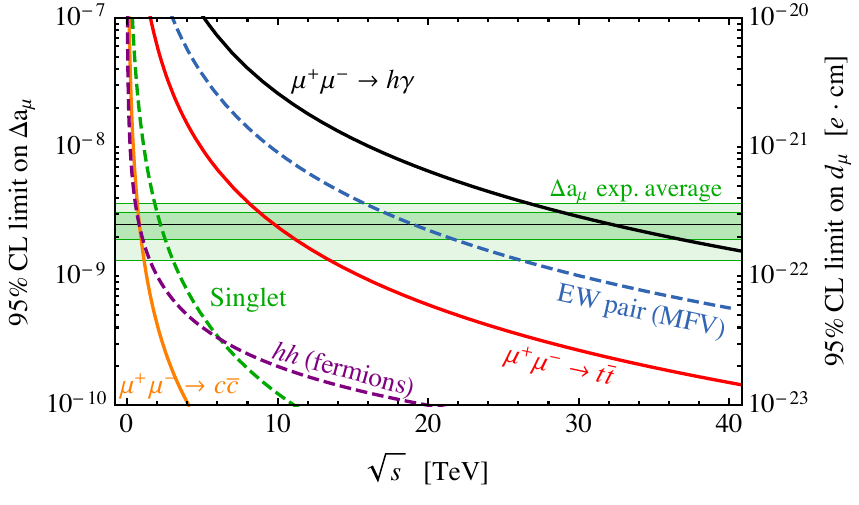}
\end{center}
\end{minipage}
\hfill
\begin{minipage}{0.46\textwidth}
\begin{center}
\includegraphics[width=\textwidth]{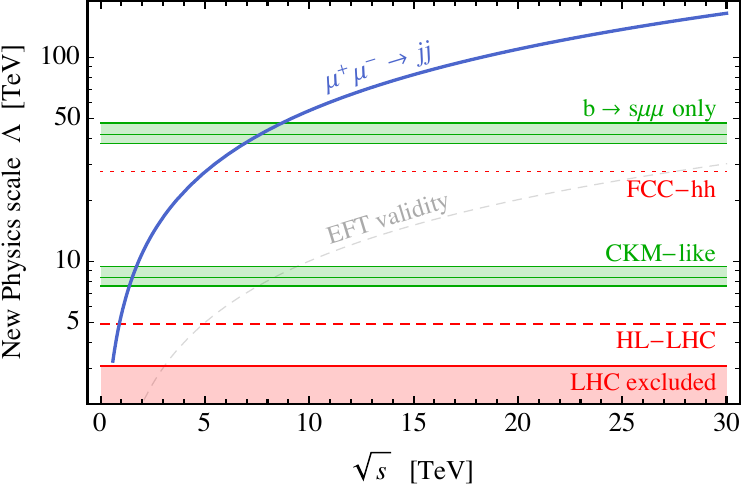}
\end{center}
\end{minipage}
\caption{Summary, from Ref.~\cite{MuonCollider:2022xlm}, of the muon collider sensitivity to putative new physics responsible for the muon anomalies. Left panel: reach on the muon $g$-2 from high-energy measurements (solid lines), and from direct searches for new particles in explicit models (dashed lines). Right panel: reach from $\mu\mu\to jj$ (solid line) on the scale $\Lambda$ of semi-leptonic interactions that can account for the $B$-anomalies.
\label{fan}}
\end{figure}

A muon collider offers excellent prospects to probe putative new physics scenarios responsible for the muon anomalies, as schematically summarized in Figure~\ref{fan}. The left panel reports the minimal muon collider energy that is needed to probe different types of new physics potentially responsible for the $g$-2 anomaly. The solid lines correspond to limits on contact interaction operators due to unspecified new physics, that contribute at the same time to the muon $g$-2 and to high-energy scattering processes. Semi-leptonic muon-charm (muon-top) interactions that can account for the $g$-$2$ discrepancy can be probed by di-jets at a $3$~TeV ($10$~TeV) muon collider, whereas a 30~TeV collider could even probe a tree-level contribution to the muon electromagnetic dipole operator directly through $\mu\mu\to h\gamma$. These sensitivity estimates are agnostic on the specific new physics model responsible for the anomaly. Explicit models typically predict light particles that can be directly discovered at the muon collider, and correlated deviations in additional observables. In the figure, dashed lines illustrate the sensitivity to three classes of models:  those featuring EW-singlet scalars or vectors, the ones including EW-charged particles in models with minimal flavour violation (MFV), and heavy lepton-like particles that mix with the muon. A complete coverage of several models is possible already at a $3$~TeV muon collider, and a collider of tens of TeV could provide a full-fledged no-lose theorem.

The right panel of Figure~\ref{fan} exemplifies instead the muon collider potential to probe explanations of the flavour anomalies, in an effective field theory description of the associated new physics. The green band labeled ``$b\to s\mu\mu$ only'' represents the scale~$\Lambda$ of the interaction operator responsible for the $R_{K^{(*)}}$ anomaly (with $1/\Lambda^2$ being the Wilson coefficient). This scenario would not be testable at the FCC-hh proton collider, but it would be within the reach of a muon collider with $7$~TeV energy or more by measuring the $\mu^+\mu^-\hspace{-4pt}\to {\rm{jets}}$ cross-section induced by the same operator. Moreover in realistic new physics models the $(bs)(\mu\mu)$ interaction is unavoidably accompanied by flavour-conserving $(bb)(\mu\mu)$ and $(ss)(\mu\mu)$ interactions with a larger Wilson coefficient, corresponding to a smaller~$\Lambda$ scale reported in the ``CKM-like'' band. In particular the band assumes a $V_{ts}$ suppression of the $(bs)$ operator relative to the operators that are diagonal in the quark flavour, as it would emerge in models with a realistic flavour structure. The new physics scale $\Lambda$ is in this case within the reach of a $3$~TeV muon collider, while it cannot be probed by the HL-LHC. Of course, these considerations hold if the new particles are heavy and the EFT description is valid. If the new physics is weakly coupled and the new states are light enough, they can be directly produced at a muon collider or a hadron collider of suitable energy. See Ref.~\cite{MuonCollider:2022xlm} for more details, for a comprehensive investigation of explicit models and for an assessment of the muon collider direct sensitivity.

The muon-related anomalies should be regarded, as of today, as a specific illustration of the generic added value for new physics exploration of a collider that employs second-generation particles. However in a few years these anomalies might turn, if confirmed, into a primary driver of particle physics research. Muon colliders offers excellent perspectives for progress on the anomalies already at $3$~TeV, with a very competitive time scale. This scenario further supports the urgency of investing in a complete muon collider design study.

\section{Electroweak radiation}\label{EWRadiation}

The novel experimental setup offered by lepton collisions at $10$~TeV energy or more outlines offers possibilities for theoretical exploration that are at once novel and speculative, yet robustly anchored to reality and to phenomenological applications. 

The muon collider will probe for the first time a new regime of EW interactions, where the scale $m_{{\rm{\textsc{w}}}}\hspace{-2pt}\sim\hspace{-2pt}100~$GeV of EW symmetry breaking plays the role of a small IR scale, relative to the much larger collision energy. This large scale separation triggers a number of novel phenomena that we collectively denote as ``EW radiation'' effects. Since they are prominent at muon collider energies, the comprehension of these phenomena is of utmost importance not only for developing a correct physical picture but also to achieve the needed accuracy of the theoretical predictions.

The EW radiation effects that the muon collider will observe, which will play a crucial in the assessment of its sensitivity to new physics, can be broadly divided in two classes. 

The first class includes the initial-state radiation of low-virtuality vector bosons. It {\it effectively} makes the muon collider a high-luminosity vector bosons collider, on top of a very high-energy lepton-lepton machine. The compelling associated physics studies described in Section~\ref{VBF} pose challenges for  fixed-order theoretical predictions and Monte Carlo event generation even at tree-level, owing to the sharp features of the Monte Carlo integrand induced by the large scale separation and the need to correctly handle QED and weak radiation at the same time, respecting EW gauge invariance. Strategies to address these challenges are available in  {\texttt{WHIZARD}}~\cite{Kilian:2007gr}, they have been recently implemented in {\texttt{MadGraph5\_aMC@NLO}}~\cite{Costantini:2020stv, Ruiz:2021tdt} and applied to several phenomenological studies in the muon collider context. Dominance of such initial-state collinear radiation will eventually require a systematic theoretical modeling in terms of EW Parton Distribution Function where multiple collinear radiation effects are resummed. First studies show that EW resummation effects can be significant at a 10~TeV muon collider~\cite{Han:2020uid}.

The second class of effects are the virtual and real emissions of soft and soft-collinear EW radiation. They affect most strongly the measurements performed at the highest energy, described in Section~\ref{HEM}, and impact the corresponding cross-section predictions at order one~\cite{Chen:2022msz}. They also give rise to novel processes such as the copious production of charged hard final states out of the neutral $\mu^+\mu^-$ initial state, and to new opportunities to detect new short distance physics by studying, for one given hard final state, different patterns of radiation emission~\cite{Chen:2022msz}. The deep connection with the sensitivity to new physics makes the study of EW radiation an inherently multidisciplinary enterprise that overcomes the traditional separation between ``SM background'' and ``BSM signal'' studies. 

At very high energies EW radiation displays similarities with QCD and QED radiation, but also remarkable differences that pose profound theoretical challenges. First, being EW symmetry broken at low energy particles with different ``EW color'' are easily distinguishable. In particular the beam particles (e.g., charged left-handed leptons) carry definite color thus violating the KLN theorem assumptions. Therefore, no cancellation takes place between virtual and real radiation contributions, regardless of the final state observable inclusiveness~\cite{Ciafaloni:2000df,Ciafaloni:2000rp}. Furthermore the EW color of the final state particles can be measured, and it must be measured for a sufficiently accurate exploration of the SM and BSM dynamics. For instance, distinguishing the top from the bottom quark or the $W$ from the $Z$ boson (or photon) is necessary to probe accurately and comprehensively new short-distance physical laws that can affect the dynamics of the different particles differently. When dealing with QCD and QED radiation only, it is sufficient instead to consider ``inclusive'' observables where QCD/QED radiation effects can be systematically accounted for and organized in well-behaved (small) corrections. The relevant observables for EW physics at high energy are on the contrary dramatically affected by EW radiation effects. Second, in analogy with QCD and unlike QED, for EW radiation the IR scale is physical. However, at variance with QCD, the theory is weakly-coupled at the IR scale, and the EW ``partons'' do not ``hadronise''. EW~showering therefore always ends at virtualities of order 100~GeV, where  heavy EW states $(t,W,Z,H)$ coexist with light SM ones, and then decay.  Having a complete and consistent description of the evolution from high virtualities where EW symmetry is restored, to the weak scale where EW is broken, to GeV scales, including also leading QED/QCD effects in all regimes is a new challenge~\cite{Han:2021kes}. 

Such a strong phenomenological motivation, and the peculiarities of the problem, stimulate work and offer a new perspective on resummation and showering techniques, or more in general trigger theoretical progress on IR physics. Fixed-order calculations will also play an important role. Indeed while the resummation of the leading logarithmic effects of radiation is mandatory at muon collider energies~\cite{Chen:2022msz,Bauer:2018xag}, subleading logarithms could perhaps be included at fixed order. Furthermore one should eventually develop a description where resummation is merged with fixed-order calculations in a exclusive way, providing the most accurate predictions in the corresponding regions of the phase space, as currently done for QCD computations. 

A significant literature on EW radiation exists, starting from the earliest works on double-logarithm resummations based on Asymptotic Dynamics~\cite{Ciafaloni:2000df,Ciafaloni:2000rp} or on the IR evolution equation~\cite{Fadin:1999bq,Melles:2000gw}. The factorization of virtual massive vector boson emissions, leading to the notion of effective vector boson is also known since long~\cite{Kane:1984bb,Dawson:1984gx,Chanowitz:1985hj,Kunszt:1987tk}. More recent progress includes resummation at the next to leading logarithm in the Soft-Collinear Effective Theory framework~\cite{Chiu:2007yn,Chiu:2007dg,Chiu:2009ft,Manohar:2014vxa,Manohar:2018kfx}, the operatorial definition of the distribution functions for EW partons~\cite{Bauer:2017isx,Fornal:2018znf,Bauer:2018xag} and the calculation of the corresponding evolution, as well as the calculation of the EW splitting functions~\cite{Chen:2016wkt} for EW showering and the proof of collinear EW emission factorization~\cite{Borel:2012by,Wulzer:2013mza,Cuomo:2019siu}. Additionally, fixed-order virtual EW logarithms are known for generic process at the $1$-loop order~\cite{Denner:2000jv,Denner:2001gw} and are implemented in {\texttt{Sherpa}}~\cite{Bothmann:2020sxm} and  {\texttt{MadGraph5\_aMC@NLO}}~\cite{Pagani:2021vyk}. Exact EW corrections at NLO are available in an automatic form for arbitrary processes in the SM, for example in {\texttt{MadGraph5\_aMC@NLO}}~\cite{Frederix:2018nkq} and in {\texttt{Sherpa+Recola}}~\cite{Biedermann:2017yoi}. Implementations of EW showering are also available through a limited set of splittings in {\texttt{Pythia 8}}~\cite{Christiansen:2014kba,Christiansen:2015jpa} and in a complete way in {\texttt{Vincia}}~\cite{Brooks:2021kji}.

While we are still far from an accurate systematic understanding of EW radiation, the present-day knowledge is sufficient to enable rapid progress in the next few years. The outcome will be an indispensable toolkit for muon collider predictions. Moreover, while we do expect that EW radiation phenomena can in principle be described by the Standard Model, they still qualify as  ``new phenomena'' until when we will be able to control the accuracy of the predictions and verify them experimentally. Such investigation is a self-standing reason of scientific interest in the muon collider project.

\newpage

\section{The path to a new generation of experiments}

The rich program enabled by colliding bunches of muons requires novel detectors and reconstruction techniques to successfully exploit the physics potential of the machine.

The main challenge to operating a detector at a muon collider is the fact that muons are unstable particles. As such, it is impossible to study the muon interactions without being exposed to decays of the muons forming the colliding beams. From the moment the collider is turned on and the muon bunches start to circulate in the accelerator complex, the products of the in-flight decays of the muon beams and the results of their interactions with beamline material, or the detectors themselves, will reach the experiments contributing to polluting the otherwise clean collision environment. The ensemble of all these particles is usually known as ``Beam Induced Backgrounds'', or BIB. The composition, flux, and energy spectra of the BIB entering a detector is closely intertwined with the design of the experimental apparatus, such as the beam optics that integrate the detectors in the accelerator complex or the presence of shielding elements, and the collision energy. However, two general features broadly characterize the BIB: it is composed of low-energy particles with a broad arrival time in the detector.

The design of an optimized detector is still in its infancy, but it is already clear that the physics goals will require a fully hermetic detector able to resolve the trajectories of the outgoing particles and their energies. While the final design might look similar to those taking data at the LHC, the technologies at the heart of the detector will have to be new. The large flux of BIB particles sets requirements on the need to withstand radiation over long periods of time, and the need to disentangle the products of the beam collisions from the particles entering the sensitive regions from uncommon directions calls for high-granularity measurements in space, time and energy. The development of these new detectors will profit from the consolidation of the successful solutions that were pioneered for example in the High Luminosity LHC upgrades, as well as brand new ideas. New solutions are being developed for use in the muon collider environment spanning from tracking detectors, calorimeters systems and dedicated muon systems. The whole effort is part of the push for the next generation of high-energy physics detectors, and new concepts targeted to the muon collider environment might end up revolutionizing other future proposed collider facilities as well.

Together with a vibrant detector development program, new techniques and ideas needs to be developed in the interpretation of the energy depositions recorded by the instrumentation. The contributions from the BIB add an incoherent source of backgrounds that affect different detector systems in different ways and that are unprecedented at other collider facilities. The extreme multiplicity of energy depositions in the tracking detectors create a complex combinatorial problem that challenges the traditional algorithms for reconstructing the trajectories of the charged particles, as these were designed for collisions where sprays of particles propagate outwards from the centre of the detector. At the same time, the potentially groundbreaking reach into the high-energy frontier will lead to strongly collimated jets of particles that need to be resolved by the calorimeter systems, while being able to subtract with precision the background contributions. The challenging environment of the muon collider offers fertile ground for the development of new techniques, from traditional algorithms to applications of artificial intelligence and machine learning, to brand new computing technologies such as quantum computers.

\addcontentsline{toc}{chapter}{References}

\bibliographystyle{report}
\bibliography{report}
\end{document}